\documentclass{emulateapj}
\usepackage{natbib}
\begin{document}

\title{Protoplanetary Disks in the Orion OMC1 Region Imaged with ALMA}

\author{J. A.  Eisner\altaffilmark{1}, J. M. Bally\altaffilmark{2},
  A. Ginsburg\altaffilmark{3}, P. D. Sheehan\altaffilmark{1}}
\altaffiltext{1}{Steward Observatory, University of Arizona, 933 North
  Cherry Avenue, Tucson, AZ 85721, USA}
\altaffiltext{2}{ Department of Astrophysical and Planetary Sciences,
  University of Colorado, UCB 389, Boulder, CO 80309, USA}
\altaffiltext{3}{ESO Headquarters, Karl-Schwarzschild-Str. 2, 85748 Garching bei München, Germany}
\email{jeisner@email.arizona.edu}


%

\begin{abstract}
We present ALMA observations of the Orion Nebula that cover the OMC1
outflow region.  Our focus in this paper is on compact emission from
protoplanetary disks.  We mosaicked a field containing
$\sim 600$ near-IR-identified young stars, around which we can search for
sub-mm emission tracing dusty disks.  Approximately 100 
sources are known proplyds identified with HST.  We detect
continuum emission at 1 mm wavelengths towards $\sim 20\%$ of the proplyd
sample, and $\sim 8$\% of the larger sample of near-IR objects.  The
noise in our maps allows 4$\sigma$ detection of objects brighter than
$\sim 1.5$ mJy, corresponding to protoplanetary disk masses larger
than 1.5 M$_{\rm J}$ (using standard assumptions about dust opacities
and gas-to-dust ratios).  None of these disks are detected in
contemporaneous CO(2-1) or C$^{18}$O(2-1)
observations, suggesting that the gas-to-dust
ratios may be substantially smaller than the canonical value of 100.
Furthermore, since dust grains may already be sequestered in large
bodies in ONC disks, the inferred masses of disk solids 
may be underestimated.  Our results suggest that the distribution of
disk masses in this region is compatible with the detection rate of
massive planets around M dwarfs, which are the dominant stellar
constituent in the ONC.
\end{abstract}

\section{Introduction}
Protoplanetary disks are the birth-sites of planetary systems, and the
mass distribution of disks relates directly to the
masses of planets that may potentially form.  The
minimum-mass solar nebula (MMSN) needed to form the planets in our
Solar System is likely between 0.01 and 0.1 M$_{\odot}$
\citep[e.g.,][]{WEID+77,DESCH07}.  The evolution of disk mass with time
provides a separate, but similar constraint.  In order to build giant
planets on timescales shorter than inferred disk lifetimes, models
require disk masses $\ga 0.01$ M$_{\odot} \approx 10$ M$_{\rm J}$
\citep[e.g.,][]{HAYASHI81,ALIBERT+05}.  Measuring the distribution and
time-evolution of disk mass around young stars provides critical
constraints on planet formation.

A widely used method \citep[e.g.,][]{BECKWITH+90}
for measuring disk masses is to observe emission 
from optically thin dust, and then use assumed grain properties
to convert observed fluxes into dust masses.  The assumed opacities
imply that most of the dust mass still resides in particles smaller
than about 1 mm.  If mass is already sequestered in 
larger bodies like planetesimals (which emit much less per
unit mass), then the assumed opacities lead to underestimated disk
masses.    An assumed gas-to-dust ratio
is then used to estimate the total (gas+dust) circumstellar mass.
While gas dominates the total mass budget, the mass and surface
density of solids are crucial for understanding core formation as well
as the potential for accretion of giant planet atmospheres \citep[since fast
core formation allows gas accretion before disk dispersal; e.g.,][]{POLLACK+96}.

At short wavelengths ($\lambda \la 10$ $\mu$m), the dust in protoplanetary
disks is optically thick even for masses $<10^{-6}$ M$_{\odot}$.  
Observations at sub-mm and mm wavelengths are necessary to measure 
optically thin dust emission, and hence to determine the total mass
of dust in the disk.  


While disk masses have been well-studied in low-density star forming
regions
\citep[e.g.,][]{AW05,AW07,ANDREWS+13,WILLIAMS+13,CARPENTER+14,ANSDELL+15},
these are not the typical birth-sites
of stars.  Most stars form in rich clusters like the Orion Nebula
\citep{LADA+91,LSM93,CARPENTER00,LL03}.  Isotopic abundances in 
our Solar System suggest that it, too, may have formed in 
a dense, Orion-like environment \citep[e.g.,][]{HD05,WG07}. 
Expanding millimeter continuum surveys to include rich clusters allows the 
determination of the frequency and evolution of massive disks in
typical star (and planet) formation environments.  

Rich clusters are relatively challenging to observe because of their
distances and high stellar density. High angular resolution
and sensitivity are required.  Only a handful of rich clusters have
been observed to date: the Orion Nebula cluster
\citep{MLL95,BALLY+98,WAW05,EC06,EISNER+08,MW09,MW10,MANN+14}, 
IC 348 \citep{CARPENTER02,CIEZA+15}, and NGC 2024 \citep{EC03,MANN+15}. 
These existing surveys have detected very few disks with $\ga
0.01$--0.1 M$_{\odot}$ of material, in part because of
limited sensitivity and areal coverage.  The large number of
non-detections implies that the bulk of the disk mass distribution has
not yet been constrained.  Indeed, stacking analysis of non-detected
cluster members suggests average disk masses approximately 10 times
smaller than the detection thresholds of previous surveys
\citep[e.g.,][]{EISNER+08}.

Here we present a new 1.3 mm wavelength interferometric survey of
the Orion Nebula cluster (ONC) with ALMA.  The ONC is the logical
first choice for early ALMA observations, since it is a young, embedded
stellar cluster comprised of hundred of stars spanning a broad mass
range.   The Trapezium region contains hundreds of stars within a several
arcminute radius, and pre-main-sequence
evolutionary models \citep[e.g.,][]{DM94} fitted to spectroscopic
and/or photometric data indicate that most stars are less than 
approximately one million years old \citep[e.g.,][]{PROSSER+94,HILLENBRAND97}.
We can therefore investigate here 
correlations of disk properties with stellar and/or environmental
properties.

\section{Observations and Data Reduction}

\subsection{ALMA Cycle 2 Observations}
We mapped a region around the OMC1 BN/KL outflow in the Orion Nebula
cluster \citep[e.g.,][]{BALLY+15b,SNELL+84b}.  
The primary goal of the program (as proposed) is to study the outflow in
molecular lines, and this will be the topic of a companion paper.
Here we focus on the compact circumstellar disks within the imaged region.
The map is comprised of 108 mosaicked pointings in the northwest
region of the OMC1 outflow and 39 pointings in the southeast region.
The fields were observed at 230 GHz frequency (Band 6), corresponding
to a wavelength of 1.3 mm.  Observations were taken between 19 July
2014 and 05 April 2015.

The angular resolution of the observations was
approximately $1''$.  At the distance to Orion, $\sim 400$ pc
\citep[e.g.,][]{SANDSTROM+07,MENTEN+07,KRAUS+07}, the linear resolution is
approximately 400 AU.  Thus, any objects with radii smaller than $\sim
200$ AU will be unresolved by these observations.  ALMA Compact Array
(ACA) observations were also obtained as part of this program in order
to map the extended outflow well, but we do not use these data here
since our focus is on the compact disks.

We observed continuum emission using four 1.875 GHz- wide bands
centered at 216, 228, 231, and 233 GHz. 
Substantial parts of the bandpass contained spectral line emission and
were flagged for the continuum data reduction. 
Averaging continuum data over all spectral windows, the effective central frequency is 225 GHz.
 
One 2 GHz window was devoted to CO(2-1) at 231 GHz, and
two other windows targeted C$^{18}$O and SiO at 218 GHz, and SO at 216 GHz.  All
spectral windows had frequency resolution of about 1 MHz, providing
velocity resolution of 1.3 km s$^{-1}$.

We adopted the reduced ALMA data products.  The 
reduction included heavy flagging to remove strong spectral lines from
the continuum, as well as the standard flux, bandpass, and gain
calibrations.  The sources used for calibration included J0607-0834,
J0541-0541, J0423-013, J0725-0054, Uranus, and Ganymede.
The $uv$ data were inverted using a Briggs robust
weighting parameter of 0.5, resulting in a synthesized beam with
dimensions $\sim 1\rlap{.}''4 \times 0\rlap{.}''8$.  The rms varies
across the field because of strong extended emission from OMC1.
Typical noise values in clean regions of the continuum image are 0.5
mJy/beam.  The noise is 5 mJy km
s$^{-1}$ in a typical spectral line channel, although in channels with
strong emission from the OMC1 outflow the rms can rise to $\sim 20$
mJy km s$^{-1}$.

\subsection{GeMS Near-IR Imaging}
\begin{figure*}
\plotone{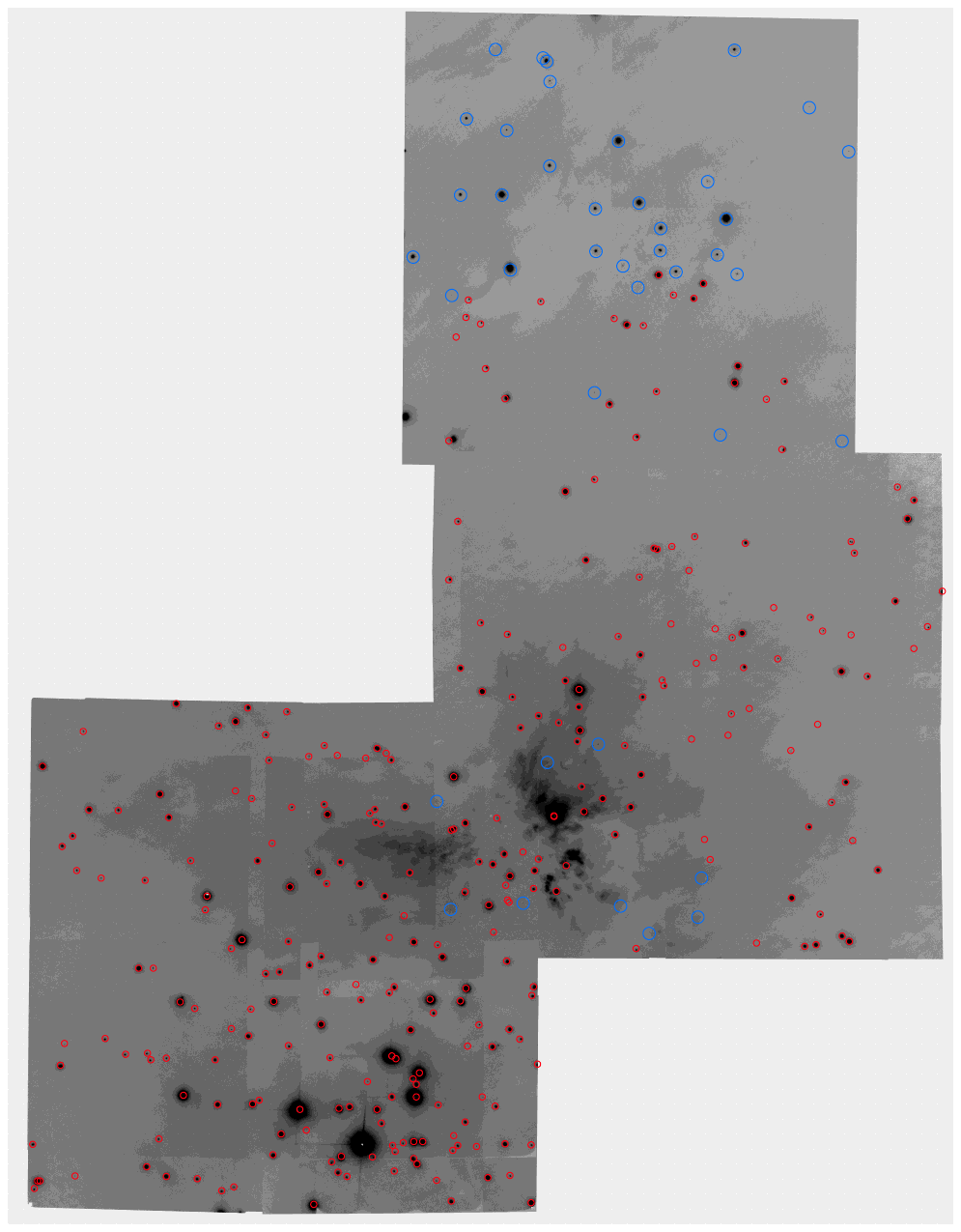}
\caption{Mosaicked image of the OMC1 region in the $K_{\rm s}$ filter,
  obtained with GeMS.  The positions of previously-known near-IR
  cluster members \citep{HC00} are indicated with small red circles.
  We detected an additional 40 stars in the image, indicated with blue
  cicles.  Some of these were outside the area covered by the survey of
  \citet{HC00}, while others can be distinguished in crowded regions
  because of the higher angular resolution of our observations.
\label{fig:gems}}
\end{figure*}

We searched for sub-mm emission toward the positions of known
near-IR sources, because detection of sub-mm continuum flux would indicate
disks \citep[as opposed to embedded spherical sources;
e.g.,][]{BECKWITH+90}.  We therefore make use of previous near-IR
observations that identified cluster members \citep{HC00}.  We
supplement this catalog with new data, because the previous
observations did not cover the complete region we mapped with ALMA.
We observed the OMC1 region 
in Orion with GeMS at the Gemini South telescope
between 30 December 2012 and 28 February 2013
\citep[see][]{BALLY+15b}.
We observed in the $K_{\rm s}$ filter, as well as narrow
Fe[II] and H$_2$ filters, producing images with angular resolutions of
$\sim 0\rlap{.}''06$.  The $K_{\rm s}$ image is shown in Figure
\ref{fig:gems}.

\begin{deluxetable}{lccc}
\tabletypesize{\scriptsize}
\tablewidth{0pt}
\tablecaption{New Near-IR Sources Detected in GeMS Imaging \label{tab:gems}}
\tablehead{\colhead{ID} & \colhead{$\alpha$} & \colhead{$\delta$} &
  \colhead{$m_{\rm K}$} \\
& (J2000) & (J2000) & }
\startdata
GeMS 1 & 5 35 12.95 & -5 22 44.34 & 14.0 \\
GeMS 2 & 5 35 12.35 & -5 22 41.35 & 14.2 \\
GeMS 3 & 5 35 15.39 & -5 22 39.93 & 14.0 \\
GeMS 4 & 5 35 13.30 & -5 22 39.32 & 13.9 \\
GeMS 5 & 5 35 14.50 & -5 22 38.78 & 14.1 \\
GeMS 6 & 5 35 12.31 & -5 22 34.19 & 14.4 \\
GeMS 7 & 5 35 15.56 & -5 22 20.13 & 13.4 \\
GeMS 8 & 5 35 14.20 & -5 22 12.99 & 13.1 \\
GeMS 9 & 5 35 13.57 & -5 22  9.65 & 13.8 \\
GeMS 10 & 5 35 10.58 & -5 21 14.08 & 13.6 \\
GeMS 11 & 5 35 12.08 & -5 21 12.95 & 14.0 \\
GeMS 12 & 5 35 13.62 & -5 21  5.22 & 13.0 \\
GeMS 13 & 5 35 15.37 & -5 20 47.37 & 14.2 \\
GeMS 14 & 5 35 13.09 & -5 20 45.92 & 14.0 \\
GeMS 15 & 5 35 11.87 & -5 20 43.51 & 11.8 \\
GeMS 16 & 5 35 12.62 & -5 20 43.07 &  9.2 \\
GeMS 17 & 5 35 14.65 & -5 20 42.70 &  7.2 \\
GeMS 18 & 5 35 13.27 & -5 20 41.99 & 12.7 \\
GeMS 19 & 5 35 15.85 & -5 20 40.34 &  8.9 \\
GeMS 20 & 5 35 12.11 & -5 20 39.95 & 10.6 \\
GeMS 21 & 5 35 13.60 & -5 20 39.30 &  9.1 \\
GeMS 22 & 5 35 12.81 & -5 20 39.16 & 11.0 \\
GeMS 23 & 5 35 12.81 & -5 20 35.10 &  8.8 \\
GeMS 24 & 5 35 12.00 & -5 20 33.40 &  7.5 \\
GeMS 25 & 5 35 13.61 & -5 20 31.51 &  8.5 \\
GeMS 26 & 5 35 13.07 & -5 20 30.42 &  8.0 \\
GeMS 27 & 5 35 15.27 & -5 20 28.99 &  9.6 \\
GeMS 28 & 5 35 14.76 & -5 20 28.99 &  7.7 \\
GeMS 29 & 5 35 12.23 & -5 20 26.52 & 12.5 \\
GeMS 30 & 5 35 14.17 & -5 20 23.65 &  9.1 \\
GeMS 31 & 5 35 10.50 & -5 20 21.03 & 13.8 \\
GeMS 32 & 5 35 13.33 & -5 20 19.09 &  8.6 \\
GeMS 33 & 5 35 14.70 & -5 20 17.13 & 11.5 \\
GeMS 34 & 5 35 15.19 & -5 20 15.01 &  9.7 \\
GeMS 35 & 5 35 10.98 & -5 20 12.94 & 13.7 \\
GeMS 36 & 5 35 14.17 & -5 20  8.17 & 11.8 \\
GeMS 37 & 5 35 14.21 & -5 20  4.50 &  7.9 \\
GeMS 38 & 5 35 14.25 & -5 20  3.84 & 11.8 \\
GeMS 39 & 5 35 11.90 & -5 20  2.42 &  9.4 \\
GeMS 40 & 5 35 14.84 & -5 20  2.29 & 14.4 \\
\enddata
\end{deluxetable}

\epsscale{0.9}

We used the IDL implementation of the {\it find} algorithm to identify
stellar sources in the $K_{\rm s}$-band image.  
We cross-referenced these with the
\citet{HC00} catalog, and kept only previously un-identified
objects.  We also visually inspected all sources, and discarded
several spurious detections in the bright central regions of the OMC1
outflow.  40 new detections remain after this process.  We performed
aperture photometry for these sources, and used previous measurements
of \citet{HC00} for stars in our field to calibrate the flux scale.
Source positions and magnitudes are listed in Table \ref{tab:gems}.

Without proper motions, the cluster membership of these
newly identified sources is uncertain.  However all objects are also
detected in narrowband filters tuned to Fe[II] in the $H$-band and
H$_2$ in the $K$-band.  Comparison of the color-magnitude relations
of new detections and previously cataloged sources in \citet{HC00}
suggest that they trace the same stellar population.  This suggests
that contamination by background or foreground objects, while
possible, is likely no worse than in the sample from \citet{HC00}.

\section{Results}

\begin{deluxetable*}{lcccccc}
\tabletypesize{\scriptsize}
\tablewidth{0pt}
\tablecaption{Fluxes and Inferred Disk Masses for ALMA-detected Sources \label{tab:detections}}
\tablehead{\colhead{ID} & \colhead{$\alpha$} & \colhead{$\delta$} &
  \colhead{$M_{\ast}$} & \colhead{$F_{\rm \lambda 1.3 mm}$} & \colhead{$F_{\rm dust}$} &
  \colhead{$M_{\rm disk}$} \\
& (J2000) & (J2000) & (M$_{\odot}$) & (mJy) & (mJy) & (M$_{\rm Jup}$)}
\startdata
HC455 &  5 35  8.93 &  -5 22 30.00 &  0.2 &  1.3 $\pm$  0.2 &  1.3 $\pm$  0.2 &  1.6 $\pm$  0.2 \\
HC592 &  5 35  9.35 &  -5 21 41.60 & -- &  5.3 $\pm$  0.4 &  5.3 $\pm$  0.4 &  6.5 $\pm$  0.5 \\
HC588 &  5 35  9.93 &  -5 21 43.40 & -- &  2.3 $\pm$  0.3 &  2.3 $\pm$  0.3 &  2.8 $\pm$  0.4 \\
102-233 &  5 35 10.13 &  -5 22 32.74 &  0.5 &  2.0 $\pm$  0.4 &  2.0 $\pm$  0.4 &  2.5 $\pm$  0.5 \\
102-021 &  5 35 10.19 &  -5 20 20.99 & -- &  5.4 $\pm$  0.5 &  5.4 $\pm$  0.5 &  6.7 $\pm$  0.6 \\
HC647 &  5 35 10.32 &  -5 21 13.10 &  0.3 & 16.3 $\pm$  1.3 & 16.3 $\pm$  1.3 & 20.1 $\pm$  1.6 \\
HC374 &  5 35 10.39 &  -5 22 59.80 & -- & 12.3 $\pm$  1.1 & 12.3 $\pm$  1.1 & 15.2 $\pm$  1.4 \\
HC421 &  5 35 10.58 &  -5 22 44.80 &  0.3 &  2.1 $\pm$  0.4 &  2.1 $\pm$  0.4 &  2.6 $\pm$  0.5 \\
106-156 &  5 35 10.58 &  -5 21 56.24 & -- & 18.6 $\pm$  1.5 & 18.6 $\pm$  1.5 & 23.0 $\pm$  1.9 \\
HC582 &  5 35 10.97 &  -5 21 46.40 & -- &  6.3 $\pm$  0.7 &  6.3 $\pm$  0.7 &  7.8 $\pm$  0.9 \\
HC415 &  5 35 11.04 &  -5 22 46.70 & -- &  9.8 $\pm$  0.9 &  9.8 $\pm$  0.9 & 12.1 $\pm$  1.1 \\
HC750 &  5 35 11.42 &  -5 21 44.60 & -- &  5.6 $\pm$  0.6 &  5.6 $\pm$  0.6 &  6.9 $\pm$  0.7 \\
HC611 &  5 35 11.77 &  -5 21 32.80 & -- &  2.7 $\pm$  0.4 &  2.7 $\pm$  0.4 &  3.3 $\pm$  0.5 \\
HC573 &  5 35 11.81 &  -5 21 49.30 & -- &  3.0 $\pm$  0.7 &  0.0 $\pm$  0.9 &  0.0 $\pm$  1.1 \\
HC672 &  5 35 11.86 &  -5 21  0.30 & -- &  9.3 $\pm$  0.8 &  9.3 $\pm$  0.8 & 11.5 $\pm$  1.0 \\
124-132 &  5 35 12.38 &  -5 21 31.39 & -- &  2.8 $\pm$  0.6 &  2.8 $\pm$  0.6 &  3.5 $\pm$  0.7 \\
GEMS23 &  5 35 12.81 &  -5 20 35.10 & -- &  1.8 $\pm$  0.3 &  1.8 $\pm$  0.3 &  2.2 $\pm$  0.4 \\
HC606 &  5 35 12.85 &  -5 21 33.90 & -- & 10.7 $\pm$  1.3 & 10.7 $\pm$  1.3 & 13.2 $\pm$  1.6 \\
HC608 &  5 35 12.89 &  -5 21 33.70 & -- &  8.3 $\pm$  1.3 &  8.3 $\pm$  1.3 & 10.2 $\pm$  1.6 \\
131-046 &  5 35 13.05 &  -5 20 45.79 & -- &  1.8 $\pm$  0.3 &  1.8 $\pm$  0.3 &  2.2 $\pm$  0.4 \\
HC682 &  5 35 13.22 &  -5 20 52.80 &  0.5 &  4.2 $\pm$  0.4 &  4.2 $\pm$  0.4 &  5.2 $\pm$  0.5 \\
132-042 &  5 35 13.24 &  -5 20 41.94 & -- &  1.5 $\pm$  0.4 &  1.5 $\pm$  0.4 &  1.9 $\pm$  0.5 \\
HC602 &  5 35 13.73 &  -5 21 35.90 & -- &  4.9 $\pm$  1.0 &  4.9 $\pm$  1.0 &  6.1 $\pm$  1.2 \\
138-207 &  5 35 13.78 &  -5 22  7.39 & -- & 25.8 $\pm$  6.8 & 25.8 $\pm$  6.8 & 31.9 $\pm$  8.4 \\
GEMS37 &  5 35 14.21 &  -5 20  4.50 & -- &  2.0 $\pm$  0.4 &  2.0 $\pm$  0.4 &  2.5 $\pm$  0.5 \\
GEMS38 &  5 35 14.25 &  -5 20  3.84 & -- &  1.8 $\pm$  0.4 &  1.8 $\pm$  0.4 &  2.2 $\pm$  0.5 \\
HC657 &  5 35 14.72 &  -5 21  6.30 & -- &  9.3 $\pm$  1.0 &  9.3 $\pm$  1.0 & 11.5 $\pm$  1.2 \\
GEMS28 &  5 35 14.76 &  -5 20 28.99 & -- & 12.0 $\pm$  1.2 & 12.0 $\pm$  1.2 & 14.8 $\pm$  1.5 \\
158-327 &  5 35 15.79 &  -5 23 26.51 &  3.0 & 17.3 $\pm$  2.8 &  5.0 $\pm$  2.8 &  6.2 $\pm$  3.5 \\
158-323 &  5 35 15.83 &  -5 23 22.59 &  0.8 & 11.8 $\pm$  2.8 &  2.0 $\pm$  2.8 &  2.5 $\pm$  3.5 \\
HC342 &  5 35 15.85 &  -5 23 11.00 &  5.0 &  6.5 $\pm$  1.2 &  6.5 $\pm$  1.2 &  8.0 $\pm$  1.5 \\
HC370 &  5 35 15.88 &  -5 23  2.00 & -- &  3.9 $\pm$  0.9 &  4.0 $\pm$  1.3 &  5.0 $\pm$  1.7 \\
163-317 &  5 35 16.27 &  -5 23 16.51 & -- & 12.1 $\pm$  1.8 &  2.0 $\pm$  1.9 &  2.5 $\pm$  2.3 \\
167-317 &  5 35 16.74 &  -5 23 16.51 &  3.0 & 21.2 $\pm$  2.3 &  3.0 $\pm$  2.3 &  3.7 $\pm$  2.8 \\
168-326 &  5 35 16.83 &  -5 23 25.91 & -- & 17.1 $\pm$  2.4 &  4.0 $\pm$  2.4 &  5.0 $\pm$  3.0 \\
170-301 &  5 35 16.95 &  -5 23  0.91 & -- &  3.6 $\pm$  0.7 &  3.0 $\pm$  0.9 &  3.7 $\pm$  1.1 \\
170-249 &  5 35 16.96 &  -5 22 48.51 &  0.4 & 12.0 $\pm$  1.4 &  9.0 $\pm$  2.1 & 11.1 $\pm$  2.5 \\
173-236 &  5 35 17.34 &  -5 22 35.81 &  3.0 & 13.6 $\pm$  1.4 & 12.0 $\pm$  2.1 & 14.8 $\pm$  2.5 \\
HC422 &  5 35 17.38 &  -5 22 45.80 & -- &  3.1 $\pm$  0.7 &  3.1 $\pm$  0.7 &  3.8 $\pm$  0.9 \\
175-251 &  5 35 17.47 &  -5 22 51.26 &  0.3 &  2.9 $\pm$  0.7 &  3.0 $\pm$  1.2 &  3.7 $\pm$  1.5 \\
178-258 &  5 35 17.84 &  -5 22 58.15 &  0.2 &  5.2 $\pm$  0.7 &  5.2 $\pm$  0.7 &  6.4 $\pm$  0.9 \\
181-247 &  5 35 18.08 &  -5 22 47.10 & -- &  3.1 $\pm$  0.6 &  5.0 $\pm$  0.8 &  6.2 $\pm$  1.0 \\
HC436 &  5 35 18.38 &  -5 22 37.50 &  0.7 &  3.6 $\pm$  0.5 &  0.0 $\pm$  0.5 &  0.0 $\pm$  0.6 \\
HC482 &  5 35 18.85 &  -5 22 23.10 &  0.2 &  8.3 $\pm$  0.8 &  8.3 $\pm$  0.8 & 10.2 $\pm$  1.0 \\
HC351 &  5 35 19.07 &  -5 23  7.50 & -- &  5.6 $\pm$  0.6 &  5.6 $\pm$  0.6 &  6.9 $\pm$  0.7 \\
191-232 &  5 35 19.13 &  -5 22 31.20 & -- &  1.8 $\pm$  0.4 &  1.8 $\pm$  0.4 &  2.2 $\pm$  0.5 \\
HC366 &  5 35 19.63 &  -5 23  3.60 &  0.1 &  4.0 $\pm$  0.5 &  4.0 $\pm$  0.5 &  4.9 $\pm$  0.6 \\
198-222 &  5 35 19.82 &  -5 22 21.55 & -- &  4.6 $\pm$  0.6 &  4.0 $\pm$  0.8 &  5.0 $\pm$  1.0 \\
202-228 &  5 35 20.15 &  -5 22 28.30 & -- &  5.4 $\pm$  0.7 &  5.4 $\pm$  0.7 &  6.7 $\pm$  0.9 \\
\enddata
\tablecomments{Stellar masses, where available, are taken from the
  literature 
\citep{HILLENBRAND97,LUHMAN+00,SLESNICK+04,HILLENBRAND+13,INGRAHAM+14}.}
\end{deluxetable*}

We searched for 1.3 mm continuum emission towards the positions of
HST-detected proplyds \citep{RRS08} and near-IR detected sources
\citep{HC00}\footnote{Most of the proplyds are also detected as
  near-IR objects.}.  
We employed a detection threshold of
4$\sigma$ above the locally-determined noise level for each optical/near-IR
source.  This threshold ensures that $\ll 1$ detection is expected
from noise fluctuations across the entire sample of known cluster
members (we do expect $>1$ noise spike above the 3$\sigma$ level).

Including the newly detected sources in our GeMS imaging, our ALMA
maps included 593 near-IR targets.  Approximately 100 of these objects
are also classified as proplyds based on HST imaging.  We detected
ALMA emission towards 21 of the known proplyds, a detection rate of
$\sim 20\%$.  The detection rate among the larger sample of near-IR
targets was somewhat lower, $\sim 8\%$, with ALMA emission seen toward
47 objects (2 proplyds in our sample are not included in the near-IR sample).

Some targets are found in regions of the ONC with bright background
emission due to nearby sources or dense regions of the molecular
cloud.  We corrected the measured fluxes by subtracting the background
emission measured in annuli around each target.  This background level
is less than 10\% of the target fluxes in all cases, so represents
only a minor correction.

\epsscale{1.0}
\begin{figure*}
\plotone{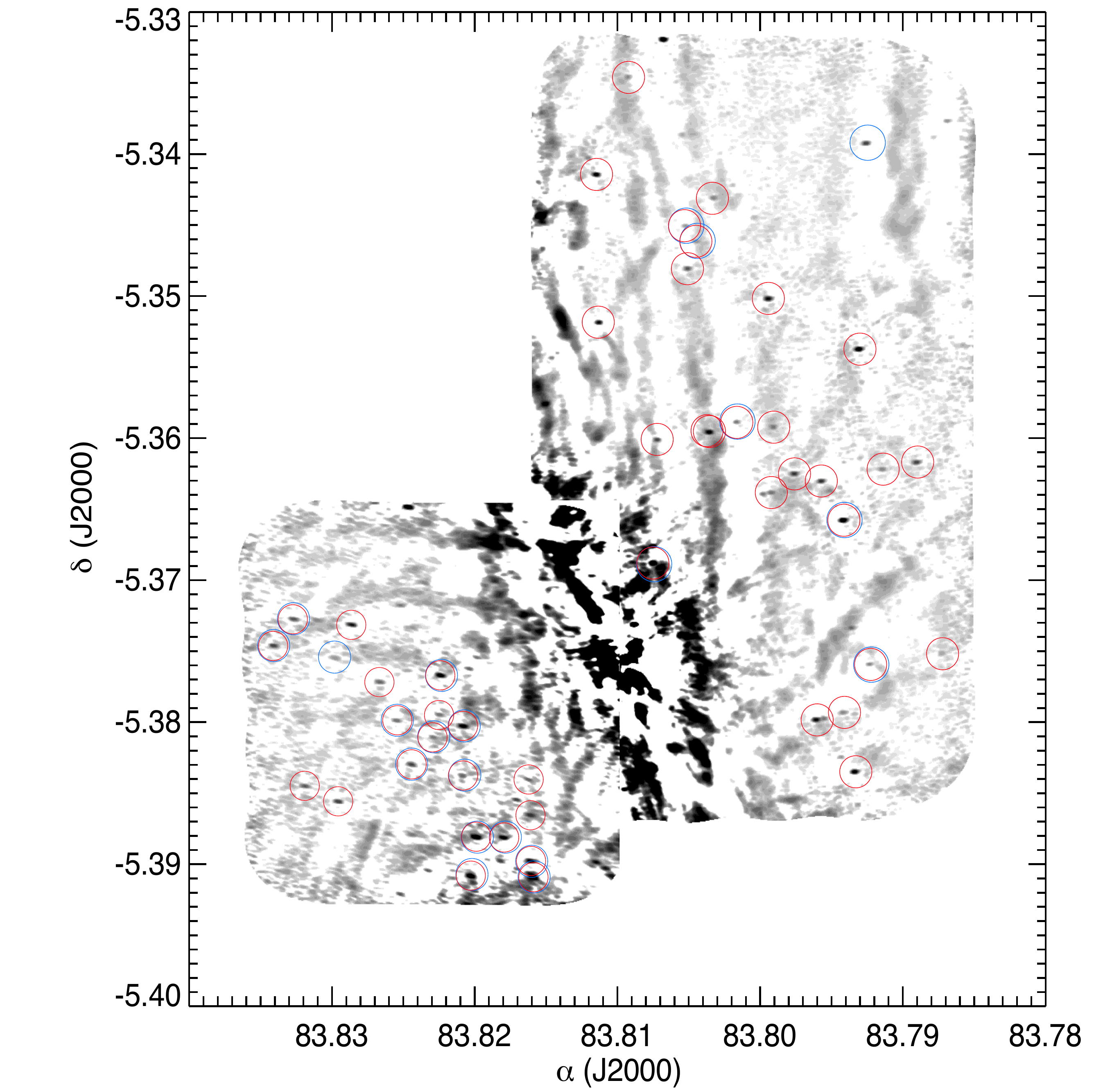}
\caption{Mosaicked image of the OMC1 region in $\lambda$1.3 mm
  continiuum emission, obtained with ALMA.  Known near-IR sources
  detected above $4\sigma$ are shown with red circles.  Blue circles
  indicate emission detected above $4\sigma$ towards known HST-imaged
  proplyds from \citet{RRS08}.  All but two of the proplyds in this
  field have known near-IR counterparts.
\label{fig:cont}}
\end{figure*}

In total, ALMA emission above the 4$\sigma$ level was seen towards 49
cluster members (Table \ref{tab:detections}; Figure \ref{fig:cont}).  
Of these, 11 were detected in previous, less
sensitive observations at a similar wavelength \citep{EISNER+08}.
Five of this sub-sample were also detected in Cycle 0 ALMA
observations at 860 $\mu$m \citep{MANN+14}, and three were detected
previously at $\lambda$3mm \citep{EC06,MLL95}.  We thus detected
millimeter-wavelength emission for the first time towards 38 stars
in the ONC.

\epsscale{1.0}
\begin{figure*}
\plotone{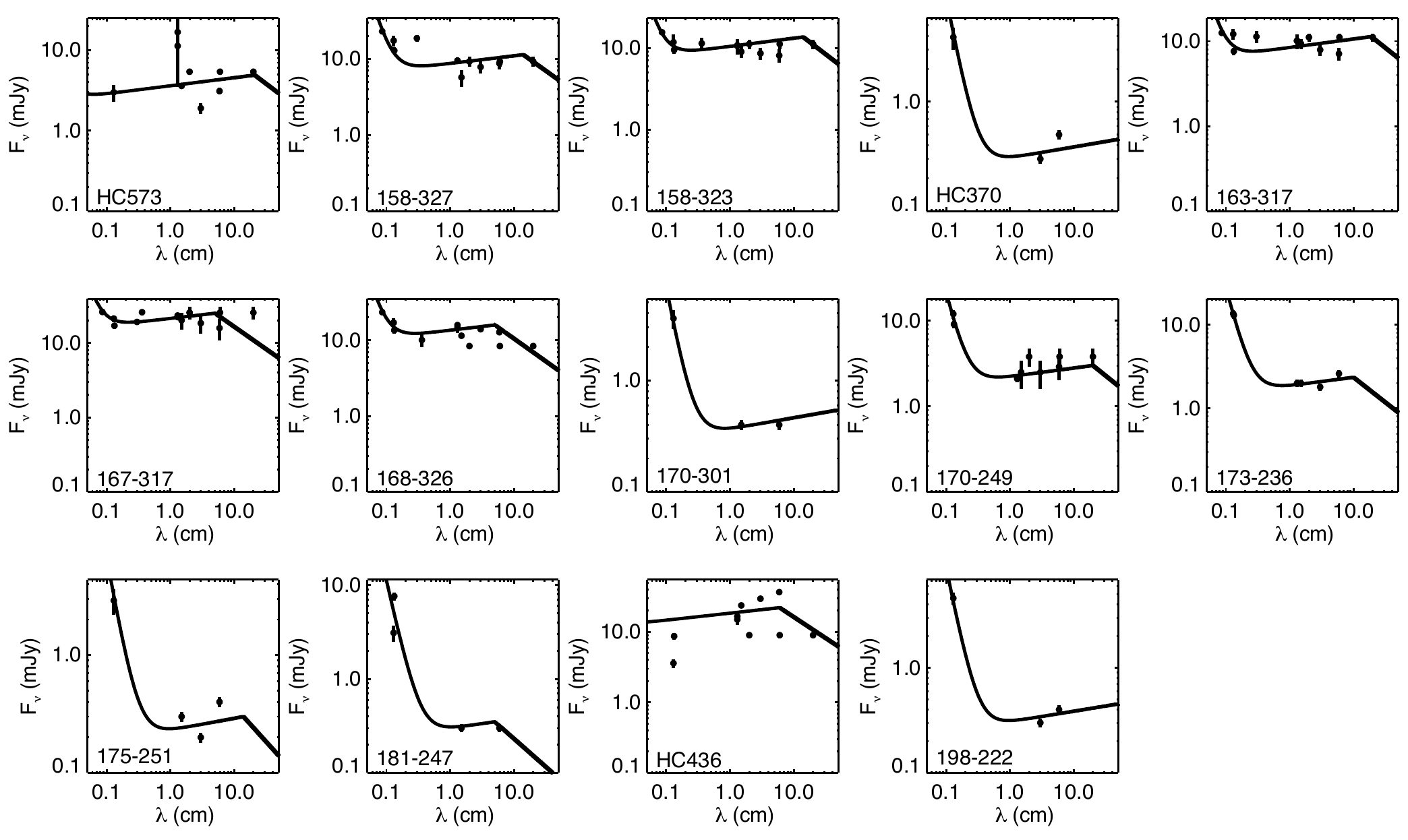}
\caption{Long-wavelength fluxes of sources detected in our ALMA
  observations that are also detected at cm wavelengths.  The solid
  curves shows models, including optically thin and thick free-free
  emission and optically thin dust emission, fitted to the data for
  each object.
  \label{fig:ff}}
\end{figure*}

Given the strong ionization field near the Trapezium stars, gas in
circumstellar disks or outflows can emit free-free emission.
Contributions of millimeter-wavelength free-free emission must be
accounted for in order to correctly determine the flux arising from
dusty disk matter.  The spectrum of optically thin free-free emission
is relatively flat compared to that of dust emission, and so
observations at cm wavelengths can be used to constrain free-free
emission \citep[e.g.,][]{EISNER+08}. 

We used previous observations of the ONC at cm wavelengths
from \citet{FELLI+93a,FELLI+93b,ZAPATA+04,FORBICH+07} to search for
free-free emission from our sample.  Nine of the sources detected in
our ALMA observations are also seen in these previous observations.
However previous surveys only constrained free-free emission at
the $\sim 1$ mJy level for most of our targets.  Given the roughly
flat spectrum of optically thin free-free emission, these measurements
can only constrain contributions at ALMA wavelengths at approximately
the same level.    Since our ALMA observations are sensitive to
emission at $<1$ mJy, we
need deeper cm-wavelength observations to constrain the
free-free emission well for our sample.

Sheehan et al. (2016) conducted a new survey of the ONC with
the JVLA, reaching sensitivities of $\la 30$ $\mu$Jy at cm wavelengths.
Their maps of 3.6 and 6 cm continuum emission cover the entire 
area of the ALMA mosaic presented here. 
In addition to the 9 targets with cm emission noted above, the deeper
cm-wavelength data reveal free-free emission from five other sources
in our sample.  The 14 objects with
cm-wavelength detections are shown in Figure \ref{fig:ff}.

For these sources, we constructed a model that includes optically
thick free-free emission past a turnover wavelength, thin free-free
emission short-ward of that turnover, and optically-thin dust with
$\beta=1$ \citep[as in][]{EISNER+08}.  Free parameters in the model
are the turnover wavelength of free-free emission, and the free-free
and dust fluxes at the observed ALMA wavelength.
Best-fit models are shown in
Figure \ref{fig:ff}.  Estimated free-free
and dust contributions to the flux at the observed ALMA wavelength
are listed in Table \ref{tab:detections}.  

The remaining objects detected in our ALMA mosaic are not seen in
cm-wavelength observations.  These non-detections imply that
the cm-wavelength flux, and hence the potential free-free contribution
at mm wavelengths, is $\la 0.03$ mJy for these sources.  Since the
observed ALMA fluxes are all $\ge 1.3$ mJy, we can be confident that
we have detected dust emission.  The $\la 0.03$ mJy uncertainties
resulting from potential low-level free-free emission are much smaller than
the uncertainties in the mm-wavelength flux measurements.

Continuum fluxes (less free-free contributions)
are converted to disk dust masses under the simple
assumption of optically thin dust:
\begin{equation}
M_{\rm dust} = \frac{S_{\rm \nu,dust} d^2} 
{\kappa_{\rm \nu,dust} B_{\nu}(T_{\rm dust})}.
\label{eq:dustmass}
\end{equation}
Here, $\nu$ is the observed frequency,
$S_{\rm \nu,dust}$ is the observed flux due to cool dust, $d$ is the distance 
to the source,
$\kappa_{\rm \nu, dust} = \kappa_0 (\nu / \nu_0)^{\beta}$ is the 
dust mass opacity,
$T_{\rm dust}$ is the dust temperature, and $B_{\nu}$ is the Planck function. 
We assume $d \approx 400$ pc,
$\kappa_0=2$ cm$^{2}$ g$^{-1}$ at 1.3 
mm, $\beta=1.0$ \citep{HILDEBRAND83,BECKWITH+90}, and $T_{\rm dust} = 20$ K
(based on the average dust temperature inferred for Taurus; Andrews \&
Williams 2005; see also the discussion in Carpenter 2002; 
Williams et al. 2005).
The dust mass can be converted into a total circumstellar mass by assuming
the canonical gas-to-dust mass ratio of 100.  For easy comparison with
the MMSN, we use these total disk masses in the figures and in Table
\ref{tab:detections}.  However we discuss potential deviations from
this assumed gas-to-dust ratio, and the implications, in \S \ref{sec:g2d}.

\begin{figure}
\plotone{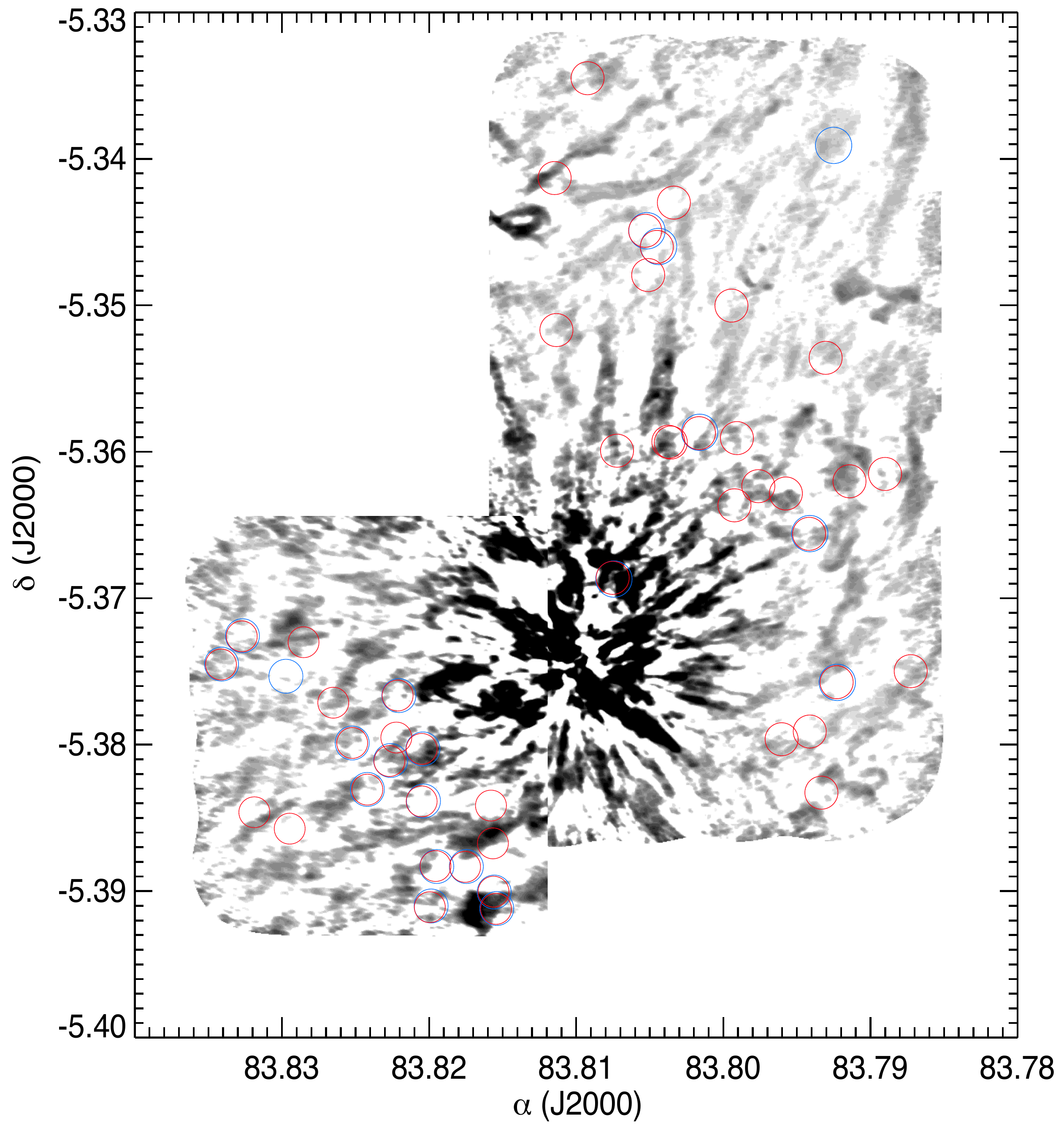}
\caption{ALMA image of the OMC1 region in line-integrated CO(2-1)
  emission.   The locations of disks detected in continuum emission
  are indicated with circles, as in Figure \ref{fig:cont}. We see no
  compact CO emission towards any of these objects.
  \label{fig:co21}}
\end{figure}

No sources are detected above the 4$\sigma$ level in CO(2-1)
emission.  The 4$\sigma$ noise level in the line-integrated map is 17
mJy km s$^{-1}$.  Indeed, for the sample of continuum detections,
where a 3$\sigma$ threshold of 12 mJy km s$^{-1}$ can be employed with 
minimal risk of false positives, we still detect no line emission from
any sources (Figure \ref{fig:co21}).
As expected, no sources are detected in C$^{18}$O emission either.

\epsscale{1.0}

 \section{Discussion}

\subsection{Disk Detection Rates \label{sec:rates}}
The detection rate among near-IR-selected sources that lie within the
ALMA mosaic is $\sim 8\%$, lower than the detection rate of $\sim
20\%$ for HST-identified proplyds (all but two of which are also included in the
near-IR selected sample).  This discrepancy may 
reflect a selection bias caused by
strong free-free emission in proplyds that lead to easier detection at
mm wavelengths (despite a low intrinsic dust disk mass).
Among our sample, 7 proplyds have strong contributions from free-free
emission to the observed mm-wavelength flux.  In contrast, only two
near-IR-selected targets (that are not also identified as proplyds) are strongly
affected by free-free emission.   Assuming these objects would not
have been detected if not for their strong free-free emission, the
discrepancy in detection rates becomes smaller.  

However some difference remains, perhaps because the near-IR sample
is more likely to include some disk-less stars, or foreground or 
background objects.
Given the selection of proplyds based on their morphologies and
observed emission from ionized gas, their status as cluster members is
indeed easier to determine.

\subsection{Disk Mass Distribution \label{sec:mdist}}
\begin{figure}
\plotone{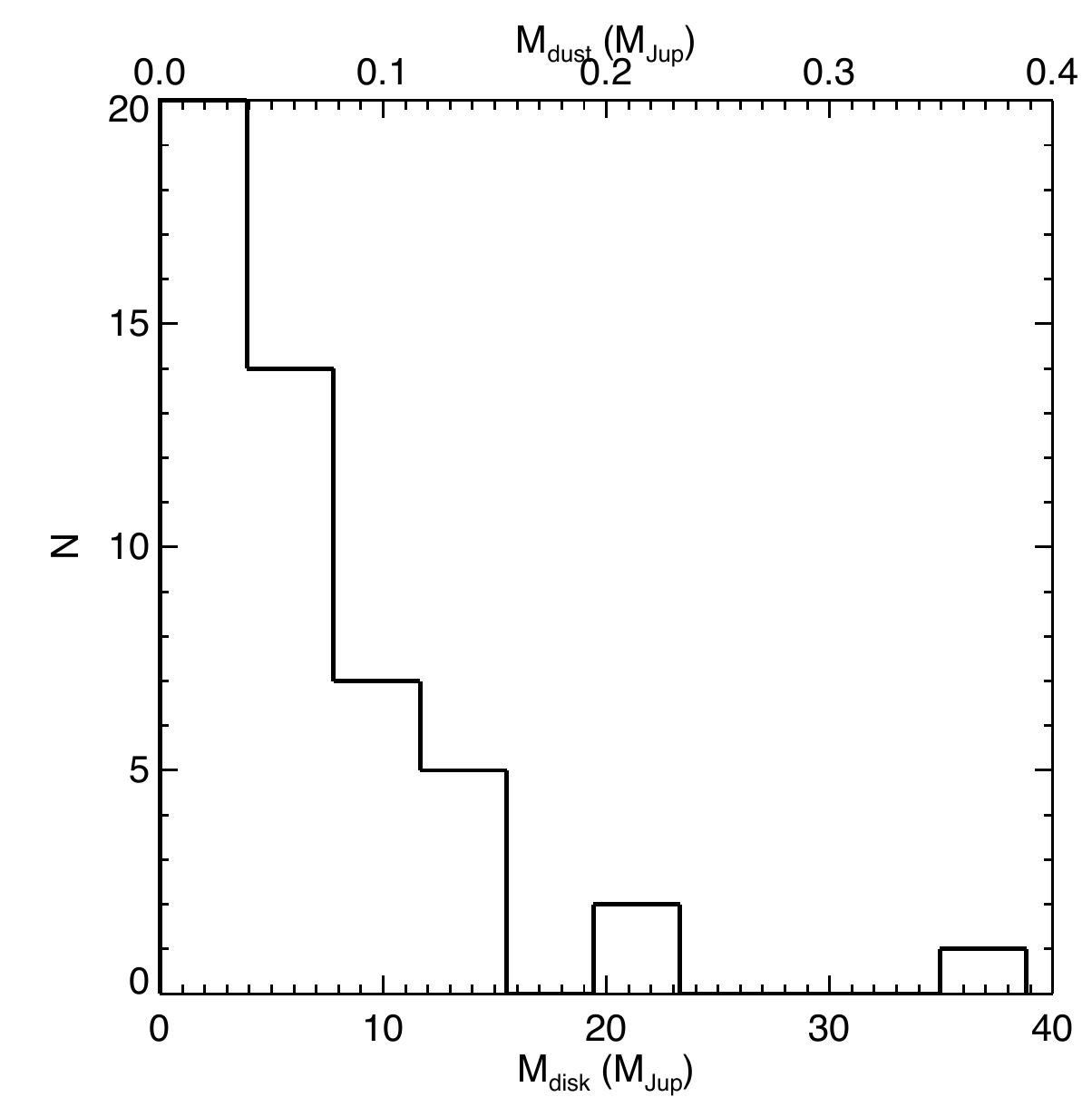}
\caption{Disk mass distribution for the sources detected with ALMA.
  We label the mass of disk dust (top axis) as well as the gas+dust
  mass (bottom axis); the latter assumes a gas-to-dust ratio of 100.
  \label{fig:mdist}}
\end{figure}

The distribution of disk masses is plotted as a histogram in Figure
\ref{fig:mdist}.  The vast majority of detections have dust disk masses
$\la 0.03$ M$_{\rm J}$ (or total gas+dust masses $\la 3$ M$_{\rm J}$
assuming a gas-to-dust ratio of 100).  The noise in the maps means
that less massive disks would probably have been undetected.  Given
that most cluster members were not detected in our observations, we
infer that the peak of the disk dust mass distribution likely lies at
masses smaller than 0.03 M$_{\rm J}$.

There are a number of objects with inferred dust+gas disk masses $\ga
0.01$ M$_{\odot}$.  From Table \ref{tab:detections}, we see that 12
sources have disk masses above this level, representing $\sim 25\%$ of
all detections, and $\sim 2\%$ of all cluster members surveyed.

\epsscale{1.0}

\subsection{Disk Mass versus Cluster Radius \label{sec:radius}}
The intense photoionizing radiation of $\theta^1$ Ori C and other
Trapezium stars can lead to photoevaporation of disks in the ONC
\citep[e.g.,][]{JOHNSTONE+98,SC01}.  One might therefore expect to see some
dependence of disk mass on cluster radius, since disks in the cluster
outskirts would be less affected by photoionizing radiation.

Intriguingly, the inner disk fraction may decrease at larger cluster
radii \citep{HILLENBRAND+98}, contrary to this expectation.  Recent
work shows no evidence for a decrease in inner disk fraction near to O
stars in the ONC or other regions \citep{RICHERT+15}.  
However inner disks are tighly
bound to their central stars and may be less affected by
photoevaporation than outer disks.

At mm/sub-mm wavelengths, where observations are sensitive to
large, massive disks, previous observations suggested a truncated disk mass
distribution within 0.03 pc of
$\theta^1$ Ori C,  whereas a broader distribution exists at
larger cluster radii \citep{MANN+14}.   This dichotomy
was interpreted as evidence that stars
within the EUV-dominated region of the Trapezium stars experience
significant disk evaporation, while disks can survive intact in the
FUV-dominated region at larger cluster radii. 

Kinematics suggest that the stellar population is unrelaxed, and
younger than the crossing time for the entire cluster
\citep[e.g.,][]{TOBIN+09}.  However, the crossing time within a
fraction of a parsec is substantially shorter than the age of the
region.  It is therefore unclear whether stars that currently lie
within 0.03 pc of $\theta^1$ Ori C have been there for a substantial
fraction of the lifetime of the cluster.  On the other hand, it may be
that the ionization front from  $\theta^1$ Ori C has turned on
recently, uncovering proplyds as it advances through the cloud.

\begin{figure}
\plotone{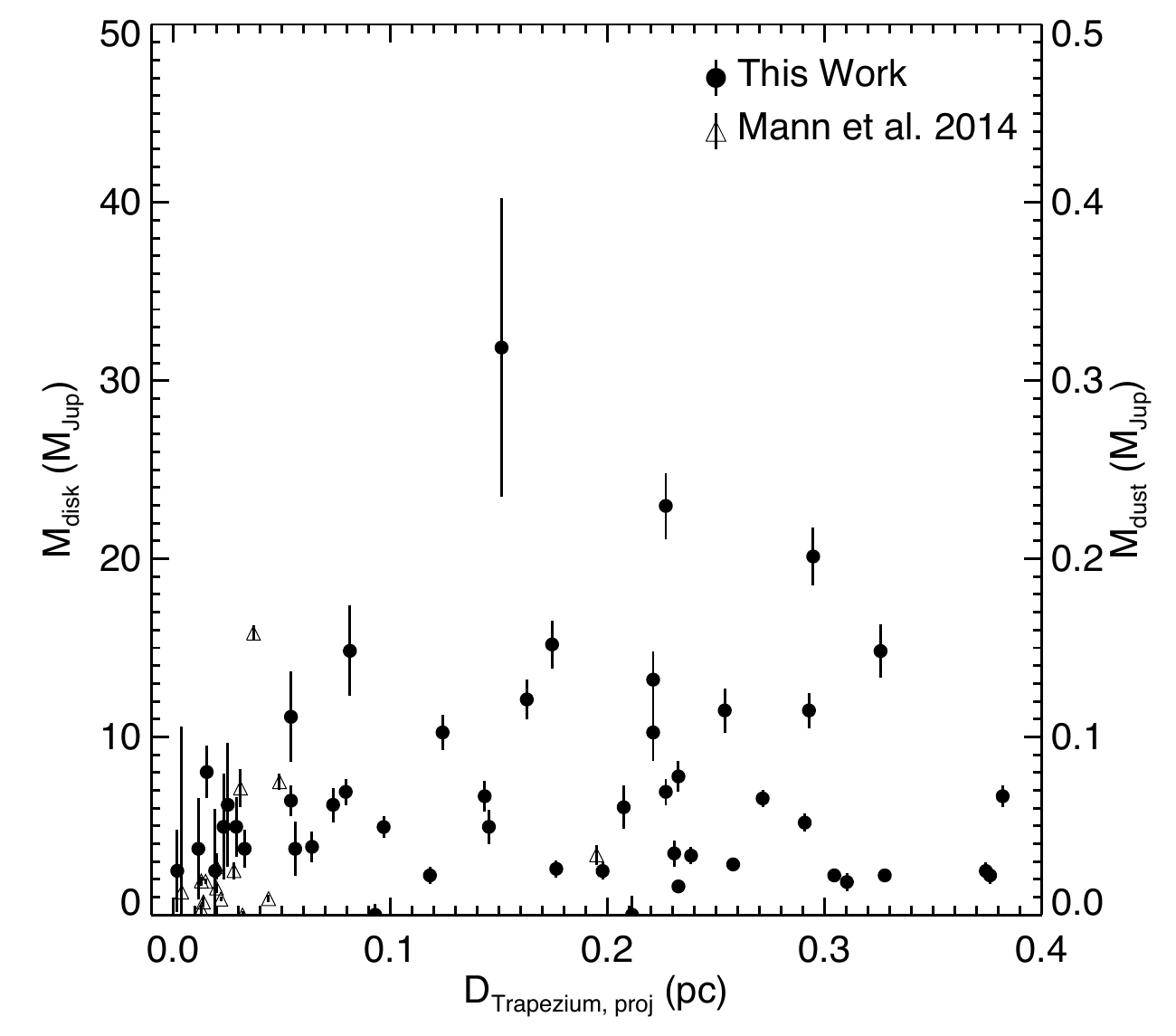}
\caption{Disk mass versus projected separation from the Trapezium
  stars, for the sources detected with ALMA.  Our measurements are
  plotted with filled circles.  We also include the measured disk
  masses for additional sources from \citet{MANN+14}, plotting them
  as open triangles.  
  Two massive disks detected at $\ga 1$ pc distance by \citet{MANN+14} are
  excluded from this plot.
  \label{fig:rdist}}
\end{figure}

Our observations support the conclusion of \citet{MANN+14}.
As seen in Figure \ref{fig:rdist}, disk masses within
about 0.03 pc range from 0 to $<10$ M$_{\rm J}$, while the range extends to
$>30$ M$_{\rm J}$ at larger distances.  This result is not
statistically significant based on our dataset. The Fisher exact test
suggests that there is a 9\% probability that a single disk mass
distribution applies within 0.03 pc and at larger radial distances.

If we include the ALMA data of \citet{MANN+14}, which are sensitive
to the same range of disk masses probed in this work, we can improve
the significance of this result.  Using both datasets, the Fisher
exact test gives only a 0.7\% probability of a single distribution,
suggesting a difference between the disk population within and beyond
$\sim 0.03$ pc of $\theta^1$Ori C significant at $\sim 3\sigma$.  Even if we
exclude the two very massive disks in \citet{MANN+14} at $\ga 1$ pc
distance, the probability of a single distribution is still only
$\sim 1\%$.

\epsscale{1.0}

\subsection{Disk Mass versus Stellar Mass \label{sec:stellar}}
Previous investigations of near-IR excess emission
showed the inner disk fraction for stars in Orion
to be largely independent of stellar age and mass, although there are
indications of a paucity of disks around very massive stars
\citep{HILLENBRAND+98,LADA+00}.  However this correlation has not been
studied in the ONC for massive disks as traced by mm/sub-mm emission.

\begin{figure}
\plotone{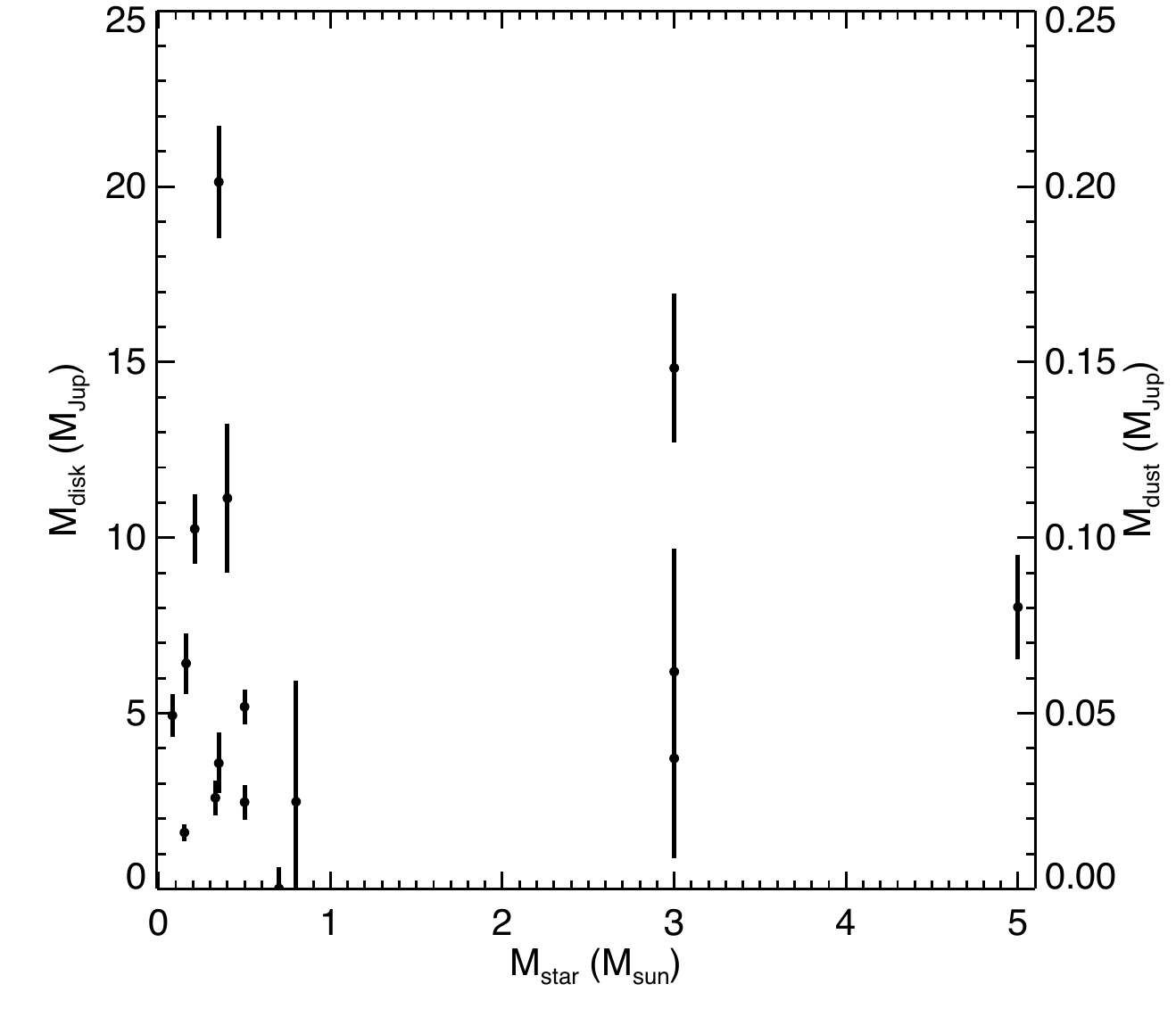}
\caption{Disk mass versus stellar mass for the subset of our sample
  with spectroscopically-determined stellar masses.
  \label{fig:massmass}}
\end{figure}

Observations of Taurus suggest a significant correlation between disk
and stellar mass \citep{ANDREWS+13}.   While we see no such
correlation in the ONC, at present the statistical significance of
this result is limited both by the small number of ALMA detections and
the small number of spectroscopically-determined stellar masses for
extincted cluster members.    To date, stellar masses have been
determined spectroscopically for less than half of the sources in
our ALMA-detected sample
\citep{HILLENBRAND97,LUHMAN+00,SLESNICK+04,HILLENBRAND+13,INGRAHAM+14}.
We plot disk mass versus stellar mass for this sub-sample in Figure
\ref{fig:massmass}.
In the near future, deeper ALMA observations will yield a larger
sample of disk masses, and near-IR spectroscopy will provide stellar
masses of embedded cluster members in the ONC.

\epsscale{1.0}

\subsection{Gas-to-Dust Ratios \label{sec:g2d}}

While we determined dust masses above, and converted them to total
gas+dust masses using an assumed gas-to-dust ratio,
one can also use spectral line data to attempt to measure gas masses
directly \citep[e.g.,][]{WB14,MIOTELLO+14}.    Models presented in
\citet{WB14} suggest that a disk with total mass of 0.01 M$_{\odot}$
should exhibit a $^{12}$CO(2-1) flux of $\sim 0.1$--5 Jy km s$^{-1}$.
While this spans a wide range \citep[indeed,][suggest using
line ratios for more accurate gas mass estimates]{WB14,MIOTELLO+14},
even the
lower limit is higher than the detection threshold in our line maps, $\sim
12$ mJy km s$^{-1}$.  Thus, we would expect that most disks in our
sample detected in 1.3 mm continuum emission with ALMA
should be detected in CO.  

The models in \citet{WB14} included a range of disk sizes, with radii
as small as 30 AU.  Since $^{12}$CO(2-1) is generally optically thick,
still smaller disks---at fixed mass---might produce lower flux.
However, inferred sizes of proplyds are typically larger than the
lower limit in \citet{WB14}.  The mean disk radius for a sample of 134
proplyds observed with {\it HST} is 35 AU \citep{VA05}.  Many
proplyds, including some within 0.03 pc of $\theta^1$Ori C, have radii
greater than 100 AU \citep{VA05,EISNER+08}.  Thus it seems unlikely
that the lack of CO detections can be explained by unusually compact
disks.

The non-detection of any disk in the CO data suggests a relative lack
of gas in these disks.  The gas-to-dust ratio is
likely much smaller than 100.  This is consistent with previous
results: \citet{WB14} inferred values of the
gas-to-dust ratio of $\la 20$ in Taurus.  In the ONC, where strong
ionization from $\theta^1$ C and the other Trapezium stars can cause
strong photoevaporative winds from circumstellar disks, we expect
still lower gas-to-dust ratios \citep[e.g.,][]{TB05b}.

Detecting gas from disks in the ONC likely requires better sensitivity
than the observations presented here.  Assuming a gas-to-dust ratio of
10, the distribution of $M_{\rm dust}$ values shown in Figure
\ref{fig:mdist} suggests that we would need sufficient sensitivity to
detect $\la 1$ M$_{\rm J}$ of gas.  Using the lower limit from
\citet{WB14}, this could represent CO(2-1) flux $\la 10$
mJy km s$^{-1}$.

One caveat is that the strong molecular emission from the Orion cloud
could obscure gas emission from disks lying in the background.  While
most proplyds appear to reside in front of the cloud, it is possible
that some gas-rich disks in the background could be missed.

There are strong detections of a few large proplyds at relatively
large cluster radii \citep[e.g.,][]{WILLIAMS+14,BALLY+15}.  Such
large, gas-rich disks would be more likely to photoevaporate if they
resided close to the Trapezium stars.  Thus, the gas content of many,
if not most, disks in the ONC, may be substantially lower than these
examples.

\subsection{Implications for Giant Planet Formation \label{sec:planets}}
The disk mass distribution shown in Figure \ref{fig:mdist} indicates that
only a small percentage of ONC cluster members have masses comparable
to or larger than the MMSN.  Adopting 10 $M_{\rm J}$ as the minimum
value of the MMSN \citep{WEID+77}, we find 12 disks above this level.
Compared to the $\sim 600$ near-IR and optically identified cluster
members surveyed, this represents a detection rate of only 2\%.

Such a low detection rate may be at odds with the known existence
of giant planets around other stars \citep[e.g.,][]{WRIGHT+11}.
Around FGK stars, the detection rate of planets more massive than
Jupiter is $\sim 10\%$, and extrapolation to larger orbital radii
suggests  that up to 20\% of FGK stars in the Galaxy may host giant
planets \citep{CUMMING+08}. The detection rate around M dwarfs
is $\sim 5$--10 times lower\footnote{The detection rate 
  for planets less massive than Neptune, however, seems to rise
  for lower-mass stars \citep[e.g.,][]{MULDERS+15}.  Since more
  massive planets put more demanding constraints on initial disk
  masses, we focus on the statistics of massive, Jupiter-like planets
  in this paper.}, $\sim 2\%$ 
\citep{CUMMING+08,JOHNSON+10}.
The distribution of stellar masses in the ONC peaks at 0.2
M$_{\odot}$ \citep{HILLENBRAND97,SLESNICK+04}, 
implying most cluster members are M dwarfs.

A giant planet detection rate near 2\% may therefore be be an
appropriate benchmark in this case.  Since $\sim 2\%$ of ONC cluster
members in our survey have disk masses similar to the MMSN, the planet
formation potential in this young region seems close to the final
outcomes seen around main-sequence M dwarfs in the field.

However the disk masses plotted in Figure \ref{fig:mdist} assume a
gas-to-dust ratio of 100.  The non-detection of any sources in CO
emission \citep[as well as previous results for young stars in
Taurus;][]{WB14} suggests that this ratio may be substantially lower.
In that case, the fraction of disks with total masses greater than 10
M$_{\rm J}$ could approach zero.

According to the approach used to derive the MMSN, this would imply
that none of the systems in the ONC can form Jupiter-mass planets.
However this conclusion can be avoided if we
re-examine a key assumption in the MMSN model.  Specifically, that
model assumes that all of the heavy elements in Jupiter and other
planets had to condense out from Solar-composition gas.  If there is
sufficient dust to build planetary cores (perhaps from an earlier
condensation event), then a much smaller amount
of gas is needed to build the planetary atmospheres.  
For example, a 30 M$_{\oplus}$ solid core, massive enough to accrete a
large gaseous atmosphere \citep[e.g.,][]{POLLACK+96}, 
requires only $\sim 1$ M$_{\rm J}$ of solar-composition
gas to become a Jupiter-like planet.  

It is the mass of disk
solids, which determine the formation timescale of a crossover-mass
core, that provides a tighter constraint on planet-formation
potential.  A more accurate definition for the minimum mass of dust
and gas needed to form a planetary system like ours may be $>0.1$ M$_{\rm
  J}$ of dust and $>1$ M$_{\rm J}$ of gas.  This argument suggests
that 2\% is the correct number for gauging giant planet formation
potential, since a gas+dust mass of 10 M$_{\rm J}$ in Figure
\ref{fig:mdist} corresponds to 0.1 M$_{\rm J}$ of dust.

Accounting for uncertain dust opacity (i.e., mass hidden in already-grown
pebbles or planetesimals), the solid content of disks in the ONC may
be still higher than inferred above.   Lower dust opacities, which
account for this hidden mass of solids, are 
physically plausible as long as they do not
push the high end of the inferred disk mass distribution into a
gravitationally unstable regime.   Assuming most stars in our sample
have masses $\la 1$
M$_{\odot}$, we would surmise that disks could not have total masses
larger than $\sim 0.1$ M$_{\odot}$ (100 M$_{\rm J}$) 
because disks more massive would become gravitationally unstable.

Figure \ref{fig:mdist} suggests that adjusting the distribution up
substantially, to account for uncertain opacity, is not
physically plausible if the gas-to-dust ratio is 100, 
since it would produce some disks with more than
100 M$_{\rm J}$ of material.  However if the gas-to-dust ratio is only
10, then such an adjustment is allowed on physical grounds.  Such an
adjustment might also produce many systems with sufficient solids to
form giant planet cores within gas disk lifetimes.

\section{Conclusions}
We presented ALMA observations of compact emission from protoplanetary
disks in the OMC1 outflow region of the Orion Nebula.  We mosaicked a
field containing
$\sim 600$ near-IR-identified young stars, of which approximately 100 
are known proplyds identified with HST.  We detected
continuum emission at 1 mm wavelengths towards $\sim 20\%$ of the proplyd
sample, and $\sim 8$\% of the larger sample of near-IR objects.  

Using standard assumptions about the dust opacities and gas-to-dust
ratios, we converted our measured fluxes into dust and dust+gas disk
masses.  The 4$\sigma$ detection threshold in our map corresponds to a
(dust+gas) disk mass of $\sim 1.5$ M$_{\rm J}$.  The fraction of disks
with more than 10 M$_{\rm J}$ is $\sim 2\%$, similar to the detection
rate of planets around M dwarfs; M dwarfs are the dominant stellar
constituent of the ONC.

None of the sources detected in continuum emission are seen in CO(2-1)
or C$^{18}$O(2-1) emission.  While computing the emission in a single
transition is highly uncertain, even the full range of existing models
predicts that disks with gas-to-dust ratio of 100 would be detected in
our observations.  Since none were, we suggest that disks in this
region have gas-to-dust ratios that are substantially lower than 100.

The disks detected here, which appear to be gas poor, still have
substantial solid content.  With a standard dust opacity assumption,
our measurements suggest that 2\% of disks have enough solids to build
Jupiter-mass cores.  However solids may already be sequestered in
larger bodies at the age of our sample, in which case our measurements
would be consistent with higher solid disk masses.  Such disks are
physically plausible, because the low gas-to-dust ratio means that
high solid masses need not lead to global gravitational instability.
Since uncertain opacities allow inferred dust disk masses to be
adjusted upwards, the potential for giant planet formation likely
exists around significantly more than 2\% of ONC cluster members.

\vspace{0.2in}
This work was supported by NSF AAG grant 1311910. 
The National Radio Astronomy Observatory is a facility of the
National Science Foundation operated under cooperative agreement by
Associated Universities, Inc. The results reported herein benefitted
from collaborations and/or information exchange within NASA's Nexus
for Exoplanet System Science (NExSS) research coordination network
sponsored by NASA’s Science Mission Directorate.

\clearpage
\LongTables

\begin{deluxetable}{lccccc}
\tabletypesize{\scriptsize}
\tablewidth{0pt}
\tablecaption{Fluxes Observed Toward Non-detected Source Positions \label{tab:nondetections}}
\tablehead{\colhead{ID} & \colhead{$\alpha$} & \colhead{$\delta$} &
\colhead{$F_{\rm \lambda 1.3 mm}$} & \colhead{Local Noise} &
  \colhead{S/N} \\
& (J2000) & (J2000) & (mJy) & (mJy) & }
\startdata
HC634 & 5 35  8.43 & -5 21 19.80 &   0.4 &   0.2 &   2.0 \\
HC382 & 5 35  8.74 & -5 22 56.70 &   0.7 &   0.4 &   1.9 \\
HC674 & 5 35  9.49 & -5 20 58.80 &   0.0 &   0.4 &   0.0 \\
HC579 & 5 35  9.53 & -5 21 48.10 &   0.2 &   0.2 &   0.7 \\
HC748 & 5 35  9.70 & -5 21 52.10 &   0.3 &   0.3 &   1.2 \\
HC627 & 5 35  9.70 & -5 21 24.90 &   0.7 &   0.2 &   3.4 \\
HC621 & 5 35  9.78 & -5 21 28.30 &   0.2 &   0.2 &   1.0 \\
HC630 & 5 35  9.90 & -5 21 22.50 &   0.1 &   0.2 &   0.5 \\
HC671 & 5 35 10.20 & -5 21  0.40 &   0.3 &   0.3 &   1.2 \\
HC553 & 5 35 10.27 & -5 21 57.20 &   0.4 &   1.3 &   0.3 \\
HC605 & 5 35 10.43 & -5 21 34.60 &   0.2 &   0.2 &   1.1 \\
HC466 & 5 35 10.45 & -5 22 27.30 &   0.2 &   0.7 &   0.3 \\
HC612 & 5 35 10.47 & -5 21 32.50 &   0.4 &   0.2 &   1.9 \\
HC570 & 5 35 10.47 & -5 21 49.60 &   0.1 &   0.4 &   0.2 \\
HC417 & 5 35 10.49 & -5 22 45.80 &   0.4 &   0.3 &   1.2 \\
HC810 & 5 35 10.50 & -5 20 21.03 &   0.0 &   0.4 &   0.1 \\
HC500 & 5 35 10.54 & -5 22 16.60 &  -0.8 &   0.6 &  -1.3 \\
HC789 & 5 35 10.58 & -5 21 14.08 &   1.3 &   1.3 &   1.0 \\
HC385 & 5 35 10.62 & -5 22 56.10 &   1.4 &   1.1 &   1.4 \\
HC363 & 5 35 10.65 & -5 23  3.40 &   1.1 &   1.1 &   1.0 \\
HC489 & 5 35 10.71 & -5 22 20.30 &  -0.1 &   0.9 &  -0.2 \\
HC392 & 5 35 10.73 & -5 22 54.50 &   0.5 &   0.6 &   0.8 \\
HC576 & 5 35 10.82 & -5 21 48.90 &   2.1 &   0.7 &   3.0 \\
HC426 & 5 35 10.85 & -5 22 40.80 &  -0.3 &   0.4 &  -0.7 \\
HC529 & 5 35 10.88 & -5 22  6.00 &  -0.1 &   0.4 &  -0.2 \\
HC416 & 5 35 10.90 & -5 22 46.40 &   2.7 &   0.9 &   2.9 \\
HC814 & 5 35 10.98 & -5 20 12.94 &  -0.0 &   0.1 &  -0.0 \\
HC473 & 5 35 10.99 & -5 22 24.80 &  -0.8 &   0.7 &  -1.0 \\
HC434 & 5 35 11.20 & -5 22 37.80 &   0.9 &   0.5 &   1.7 \\
HC515 & 5 35 11.21 & -5 22 10.80 &  -0.0 &   0.3 &  -0.1 \\
HC666 & 5 35 11.29 & -5 21  3.10 &   0.1 &   0.2 &   0.4 \\
HC643 & 5 35 11.32 & -5 21 15.60 &   0.4 &   0.2 &   2.4 \\
HC559 & 5 35 11.37 & -5 21 54.00 &   0.4 &   0.4 &   0.9 \\
HC656 & 5 35 11.51 & -5 21  6.40 &   0.4 &   0.2 &   2.1 \\
HC404 & 5 35 11.63 & -5 22 51.70 &   0.7 &   0.8 &   0.9 \\
HC709 & 5 35 11.63 & -5 22 46.10 &  -0.4 &   0.6 &  -0.6 \\
HC539 & 5 35 11.72 & -5 22  3.10 &   1.2 &   0.7 &   1.8 \\
HC555 & 5 35 11.79 & -5 21 55.60 &   0.6 &   0.7 &   0.8 \\
HC794 & 5 35 11.87 & -5 20 43.51 &   0.4 &   0.2 &   2.6 \\
HC818 & 5 35 11.90 & -5 20  2.42 &   0.1 &   0.1 &   0.9 \\
HC664 & 5 35 11.90 & -5 21  3.40 &   1.0 &   0.8 &   1.3 \\
HC708 & 5 35 11.92 & -5 22 50.90 &   0.6 &   0.9 &   0.7 \\
HC569 & 5 35 11.93 & -5 21 50.10 &   1.4 &   0.8 &   1.7 \\
HC536 & 5 35 11.94 & -5 22  4.10 &   0.7 &   1.0 &   0.7 \\
HC772 & 5 35 11.98 & -5 22  8.00 &   1.4 &   1.0 &   1.4 \\
HC395 & 5 35 11.98 & -5 22 54.20 &   0.8 &   1.2 &   0.7 \\
HC803 & 5 35 12.00 & -5 20 33.40 &  -0.1 &   0.1 &  -0.9 \\
HC790 & 5 35 12.08 & -5 21 12.95 &   0.1 &   0.3 &   0.3 \\
HC799 & 5 35 12.11 & -5 20 39.95 &   0.8 &   0.2 &   3.4 \\
HC746 & 5 35 12.14 & -5 21 48.50 &  -0.6 &   0.7 &  -0.9 \\
HC747 & 5 35 12.16 & -5 21 53.80 &  -1.3 &   0.7 &  -1.9 \\
HC751 & 5 35 12.20 & -5 22 30.80 &   0.2 &   1.3 &   0.1 \\
HC808 & 5 35 12.23 & -5 20 26.52 &  -0.0 &   0.1 &  -0.4 \\
HC752 & 5 35 12.27 & -5 22 27.10 &  -0.8 &   1.7 &  -0.5 \\
HC698 & 5 35 12.28 & -5 20 45.20 &   0.5 &   0.2 &   2.7 \\
HC785 & 5 35 12.31 & -5 22 34.19 &   1.1 &   1.6 &   0.7 \\
HC781 & 5 35 12.35 & -5 22 41.35 &   1.9 &   1.8 &   1.1 \\
HC762 & 5 35 12.37 & -5 21 54.80 &  -0.8 &   1.2 &  -0.6 \\
HC692 & 5 35 12.40 & -5 20 47.90 &  -0.0 &   0.2 &  -0.2 \\
HC753 & 5 35 12.43 & -5 22  8.70 &  -1.6 &   1.8 &  -0.9 \\
HC761 & 5 35 12.46 & -5 21 37.80 &   0.5 &   1.3 &   0.4 \\
HC368 & 5 35 12.58 & -5 23  2.00 &  -2.1 &   5.0 &  -0.4 \\
HC795 & 5 35 12.62 & -5 20 43.07 &  -0.3 &   0.2 &  -1.1 \\
HC696 & 5 35 12.65 & -5 20 47.30 &  -0.2 &   0.3 &  -0.8 \\
HC609 & 5 35 12.67 & -5 21 33.40 &   1.1 &   1.4 &   0.8 \\
HC770 & 5 35 12.68 & -5 21 47.60 &   3.6 &   1.5 &   2.4 \\
HC712 & 5 35 12.77 & -5 21 58.90 &   4.3 &   1.6 &   2.7 \\
HC706 & 5 35 12.79 & -5 21 57.90 &   4.0 &   1.6 &   2.5 \\
HC801 & 5 35 12.81 & -5 20 39.16 &   0.0 &   0.4 &   0.1 \\
HC699 & 5 35 12.83 & -5 20 43.60 &  -0.0 &   0.3 &  -0.1 \\
HC659 & 5 35 12.86 & -5 21  5.00 &   0.3 &   0.4 &   0.8 \\
HC780 & 5 35 12.95 & -5 22 44.34 &   1.6 &   6.0 &   0.3 \\
HC683 & 5 35 13.02 & -5 20 52.90 &   0.2 &   0.4 &   0.5 \\
HC544 & 5 35 13.03 & -5 22  1.00 &   1.1 &   2.0 &   0.6 \\
HC505 & 5 35 13.05 & -5 22 15.20 &   4.3 &   5.4 &   0.8 \\
HC562 & 5 35 13.06 & -5 21 53.20 &   4.4 &   1.9 &   2.3 \\
HC599 & 5 35 13.07 & -5 21 39.00 &   0.8 &   1.3 &   0.6 \\
HC805 & 5 35 13.07 & -5 20 30.42 &   0.1 &   0.3 &   0.4 \\
HC402 & 5 35 13.09 & -5 22 53.20 &  27.2 &   8.9 &   3.1 \\
HC707 & 5 35 13.11 & -5 22 47.10 &  11.9 &   6.2 &   1.9 \\
HC648 & 5 35 13.11 & -5 21 13.40 &   0.8 &   0.4 &   1.9 \\
HC487 & 5 35 13.18 & -5 22 21.20 &  18.5 &   6.3 &   2.9 \\
HC523 & 5 35 13.25 & -5 22  9.90 &   0.5 &   2.8 &   0.2 \\
HC380 & 5 35 13.29 & -5 22 57.90 &  -1.8 &   9.2 &  -0.2 \\
HC783 & 5 35 13.30 & -5 22 39.32 &   3.3 &  10.8 &   0.3 \\
HC811 & 5 35 13.33 & -5 20 19.09 &   0.9 &   0.4 &   2.5 \\
HC572 & 5 35 13.33 & -5 21 49.90 &  -2.1 &   1.3 &  -1.6 \\
HC471 & 5 35 13.37 & -5 22 26.20 &   5.2 &  11.1 &   0.5 \\
HC685 & 5 35 13.38 & -5 20 51.60 &   0.4 &   0.4 &   0.9 \\
HC653 & 5 35 13.44 & -5 21  7.40 &   0.1 &   0.3 &   0.2 \\
135-220 & 5 35 13.51 & -5 22 19.49 &  15.4 &   8.7 &   1.8 \\
HC360 & 5 35 13.53 & -5 23  4.50 &  -7.7 &   8.5 &  -0.9 \\
HC788 & 5 35 13.57 & -5 22  9.65 &   6.3 &   5.3 &   1.2 \\
HC800 & 5 35 13.60 & -5 20 39.30 &   0.8 &   0.3 &   2.3 \\
HC804 & 5 35 13.61 & -5 20 31.51 &   0.5 &   0.4 &   1.2 \\
HC633 & 5 35 13.62 & -5 21 21.10 &  -0.1 &   0.4 &  -0.3 \\
HC791 & 5 35 13.62 & -5 21  5.22 &  -0.2 &   0.3 &  -0.5 \\
HC483 & 5 35 13.75 & -5 22 22.00 &  11.2 &  21.0 &   0.5 \\
HC499 & 5 35 13.78 & -5 22 17.40 &   8.2 &  14.0 &   0.6 \\
HC541 & 5 35 13.81 & -5 22  2.80 &  17.6 &   4.8 &   3.7 \\
HC548 & 5 35 13.81 & -5 21 59.60 &  -0.5 &   4.9 &  -0.1 \\
HC525 & 5 35 13.83 & -5 22  9.10 &   6.6 &   7.8 &   0.9 \\
HC451 & 5 35 13.97 & -5 22 31.90 & -13.5 &  72.0 &  -0.2 \\
HC629 & 5 35 13.98 & -5 21 23.30 &   0.9 &   0.4 &   2.3 \\
HC552 & 5 35 13.98 & -5 21 58.00 &   5.7 &   4.5 &   1.3 \\
HC760 & 5 35 14.01 & -5 21 51.90 &   6.3 &   3.4 &   1.8 \\
HC704 & 5 35 14.06 & -5 22  5.70 &   3.4 &   5.3 &   0.7 \\
HC705 & 5 35 14.09 & -5 22 23.00 & 118.7 &  41.9 &   2.8 \\
HC438 & 5 35 14.09 & -5 22 36.60 & 252.2 &  75.7 &   3.3 \\
HC779 & 5 35 14.12 & -5 22 22.77 & 142.8 &  40.4 &   3.5 \\
142-301 & 5 35 14.15 & -5 23  0.91 &  30.3 &  10.7 &   2.8 \\
HC815 & 5 35 14.17 & -5 20  8.17 &   0.5 &   0.5 &   1.1 \\
HC809 & 5 35 14.17 & -5 20 23.65 &   0.5 &   0.4 &   1.5 \\
HC787 & 5 35 14.20 & -5 22 12.99 &  19.1 &  10.9 &   1.8 \\
HC690 & 5 35 14.28 & -5 20 48.50 &   0.5 &   0.3 &   1.5 \\
HC361 & 5 35 14.29 & -5 23  4.30 &  18.9 &  11.2 &   1.7 \\
HC458 & 5 35 14.31 & -5 22 30.70 & 224.8 & 179.8 &   1.3 \\
HC537 & 5 35 14.31 & -5 22  4.40 &  16.3 &  11.1 &   1.5 \\
HC345 & 5 35 14.32 & -5 23  8.30 &   0.0 &   1.8 &   0.0 \\
HC448 & 5 35 14.36 & -5 22 32.80 &  46.4 & 140.1 &   0.3 \\
HC399 & 5 35 14.37 & -5 22 54.10 &  28.5 &   8.9 &   3.2 \\
HC439 & 5 35 14.37 & -5 22 36.10 & 123.5 & 151.6 &   0.8 \\
HC391 & 5 35 14.39 & -5 22 55.70 &  15.0 &   7.9 &   1.9 \\
HC300 & 5 35 14.40 & -5 23 23.10 &   0.9 &   2.6 &   0.4 \\
HC784 & 5 35 14.50 & -5 22 38.78 & -30.0 &  46.5 &  -0.6 \\
HC759 & 5 35 14.50 & -5 22 29.40 & 169.6 & 143.8 &   1.2 \\
HC530 & 5 35 14.53 & -5 22  6.60 &  -6.5 &  13.2 &  -0.5 \\
HC364 & 5 35 14.54 & -5 23  3.70 &   0.4 &   1.8 &   0.2 \\
146-201 & 5 35 14.61 & -5 22  0.94 & -12.7 &  11.4 &  -1.1 \\
HC545 & 5 35 14.63 & -5 22  1.00 &  -9.0 &  12.0 &  -0.7 \\
HC796 & 5 35 14.65 & -5 20 42.70 &  -0.1 &   0.7 &  -0.2 \\
HC443 & 5 35 14.66 & -5 22 33.80 &  55.3 & 138.7 &   0.4 \\
HC276 & 5 35 14.66 & -5 23 28.70 &  -1.3 &   2.0 &  -0.6 \\
HC369 & 5 35 14.67 & -5 23  1.90 & -12.7 &   6.4 &  -2.0 \\
HC756 & 5 35 14.67 & -5 22 38.60 &  34.4 &  49.1 &   0.7 \\
HC757 & 5 35 14.69 & -5 22 38.20 & -18.3 &  78.9 &  -0.2 \\
HC575 & 5 35 14.69 & -5 21 49.50 &  -1.6 &   3.7 &  -0.4 \\
HC411 & 5 35 14.70 & -5 22 49.40 &  15.5 &   8.9 &   1.7 \\
HC812 & 5 35 14.70 & -5 20 17.13 &   0.8 &   0.3 &   2.3 \\
HC755 & 5 35 14.71 & -5 22 35.50 &  29.1 & 131.2 &   0.2 \\
HC302 & 5 35 14.72 & -5 23 22.90 &   9.1 &   2.9 &   3.1 \\
HC465 & 5 35 14.73 & -5 22 29.80 & 101.2 & 139.9 &   0.7 \\
148-305 & 5 35 14.80 & -5 23  4.76 &  -0.6 &   7.0 &  -0.1 \\
HC773 & 5 35 14.82 & -5 22 23.20 &  -1.5 &  50.4 &  -0.0 \\
HC324 & 5 35 14.84 & -5 23 16.00 &   0.9 &   2.2 &   0.4 \\
HC819 & 5 35 14.84 & -5 20  2.29 &  -0.1 &   0.4 &  -0.3 \\
HC771 & 5 35 14.86 & -5 22 44.10 &  -1.4 &  11.1 &  -0.1 \\
HC453 & 5 35 14.87 & -5 22 31.70 & -12.8 & 150.5 &  -0.1 \\
HC714 & 5 35 14.88 & -5 23  5.10 &  -1.0 &   1.9 &  -0.5 \\
149-329 & 5 35 14.92 & -5 23 29.05 &   0.3 &   1.9 &   0.1 \\
HC431 & 5 35 14.92 & -5 22 39.10 & -12.1 &  22.8 &  -0.5 \\
HC673 & 5 35 14.96 & -5 21  0.80 &   0.6 &   1.5 &   0.4 \\
150-147 & 5 35 15.00 & -5 21 47.34 &  13.6 &   5.5 &   2.5 \\
HC546 & 5 35 15.00 & -5 22  0.00 &   4.3 &  46.4 &   0.1 \\
HC334 & 5 35 15.00 & -5 23 14.30 &   2.2 &   1.6 &   1.4 \\
HC684 & 5 35 15.02 & -5 20 52.60 &   1.0 &   0.6 &   1.8 \\
HC581 & 5 35 15.02 & -5 21 47.40 &  14.3 &   5.5 &   2.6 \\
150-231 & 5 35 15.02 & -5 22 31.11 & -26.6 &  26.1 &  -1.0 \\
HC456 & 5 35 15.04 & -5 22 31.20 &   8.0 &  22.4 &   0.4 \\
HC373 & 5 35 15.04 & -5 23  1.10 &   3.4 &   2.0 &   1.7 \\
HC298 & 5 35 15.07 & -5 23 23.40 &  -1.8 &   1.8 &  -1.0 \\
HC694 & 5 35 15.17 & -5 20 48.20 &   0.4 &   1.3 &   0.3 \\
HC359 & 5 35 15.18 & -5 23  5.00 &   5.8 &   3.9 &   1.5 \\
HC813 & 5 35 15.19 & -5 20 15.01 &   0.0 &   0.7 &   0.0 \\
152-319 & 5 35 15.20 & -5 23 18.81 &   2.8 &   1.6 &   1.7 \\
HC687 & 5 35 15.20 & -5 20 51.40 &  -0.4 &   1.0 &  -0.4 \\
HC398 & 5 35 15.20 & -5 22 54.40 &  -1.1 &   1.8 &  -0.6 \\
HC478 & 5 35 15.21 & -5 22 24.10 &  10.3 &   9.2 &   1.1 \\
HC437 & 5 35 15.21 & -5 22 36.70 &   7.3 &   8.1 &   0.9 \\
HC806 & 5 35 15.27 & -5 20 28.99 &   1.0 &   1.0 &   1.0 \\
HC386 & 5 35 15.27 & -5 22 56.70 &   2.6 &   1.5 &   1.7 \\
HC558 & 5 35 15.27 & -5 21 55.70 &  12.5 &  11.1 &   1.1 \\
HC299 & 5 35 15.30 & -5 23 23.20 &  -0.1 &   1.3 &  -0.1 \\
HC624 & 5 35 15.30 & -5 21 28.80 &   2.4 &   2.1 &   1.1 \\
HC739 & 5 35 15.32 & -5 20 55.00 &  -0.8 &   0.9 &  -0.8 \\
154-324 & 5 35 15.35 & -5 23 24.11 &  -0.2 &   1.3 &  -0.2 \\
HC476 & 5 35 15.35 & -5 22 25.20 &   8.0 &   6.2 &   1.3 \\
HC310 & 5 35 15.35 & -5 23 21.40 &   1.4 &   1.5 &   0.9 \\
HC504 & 5 35 15.35 & -5 22 15.60 &  -3.4 &  12.3 &  -0.3 \\
HC792 & 5 35 15.37 & -5 20 47.37 &  -1.3 &   1.4 &  -0.9 \\
154-225 & 5 35 15.37 & -5 22 25.35 &   8.0 &   5.8 &   1.4 \\
154-240 & 5 35 15.38 & -5 22 39.85 &  25.7 &  14.1 &   1.8 \\
HC600 & 5 35 15.41 & -5 21 39.50 &   4.4 &   2.4 &   1.8 \\
HC646 & 5 35 15.41 & -5 21 14.00 &   1.3 &   1.9 &   0.6 \\
HC413 & 5 35 15.49 & -5 22 48.60 &   4.2 &   2.5 &   1.7 \\
HC327 & 5 35 15.54 & -5 23 15.80 &   0.7 &   1.3 &   0.5 \\
HC419 & 5 35 15.55 & -5 22 46.40 &   9.4 &  11.6 &   0.8 \\
HC274 & 5 35 15.56 & -5 23 29.60 &   3.0 &   2.7 &   1.1 \\
HC786 & 5 35 15.56 & -5 22 20.13 &  17.5 &  21.8 &   0.8 \\
HC378 & 5 35 15.60 & -5 22 58.90 &   4.4 &   4.7 &   0.9 \\
HC389 & 5 35 15.64 & -5 22 56.40 &   2.9 &   4.7 &   0.6 \\
157-323 & 5 35 15.72 & -5 23 22.59 &   2.5 &   2.8 &   0.9 \\
HC344 & 5 35 15.77 & -5 23  9.90 &   2.2 &   1.2 &   1.8 \\
HC598 & 5 35 15.77 & -5 21 39.80 &   2.2 &   2.2 &   1.0 \\
HC336 & 5 35 15.81 & -5 23 14.30 &   0.4 &   1.3 &   0.3 \\
158-326 & 5 35 15.81 & -5 23 25.51 &  10.7 &   2.9 &   3.7 \\
HC340 & 5 35 15.81 & -5 23 12.00 &   1.7 &   1.2 &   1.4 \\
HC291 & 5 35 15.84 & -5 23 25.60 &  10.7 &   2.8 &   3.8 \\
HC420 & 5 35 15.84 & -5 22 45.90 &   3.0 &   2.8 &   1.1 \\
HC447 & 5 35 15.89 & -5 22 33.20 &   4.6 &   3.2 &   1.4 \\
159-221 & 5 35 15.93 & -5 22 21.05 &   9.4 &   4.1 &   2.3 \\
HC490 & 5 35 15.95 & -5 22 21.10 &   9.6 &   4.2 &   2.3 \\
HC769 & 5 35 15.96 & -5 22 41.10 &   6.2 &   3.3 &   1.9 \\
HC304 & 5 35 15.97 & -5 23 22.70 &   0.9 &   2.8 &   0.3 \\
HC565 & 5 35 16.01 & -5 21 53.10 &  -1.9 &  11.7 &  -0.2 \\
161-324 & 5 35 16.05 & -5 23 24.35 &   3.5 &   2.9 &   1.2 \\
HC350 & 5 35 16.06 & -5 23  7.30 &   4.6 &   1.2 &   3.9 \\
HC296 & 5 35 16.07 & -5 23 24.30 &   3.5 &   2.8 &   1.2 \\
161-328 & 5 35 16.07 & -5 23 27.81 &   3.4 &   2.7 &   1.3 \\
HC401 & 5 35 16.08 & -5 22 54.10 &  -1.5 &   1.3 &  -1.2 \\
161-314 & 5 35 16.10 & -5 23 14.05 &   1.1 &   1.8 &   0.6 \\
HC303 & 5 35 16.10 & -5 23 23.20 &  -0.3 &   2.7 &  -0.1 \\
HC354 & 5 35 16.11 & -5 23  6.80 &   4.7 &   1.2 &   4.0 \\
HC511 & 5 35 16.12 & -5 22 12.50 &  -4.1 &   5.3 &  -0.8 \\
HC393 & 5 35 16.14 & -5 22 55.20 &  -0.1 &   1.2 &  -0.0 \\
HC768 & 5 35 16.14 & -5 22 45.10 &  -0.2 &   2.4 &  -0.1 \\
HC520 & 5 35 16.18 & -5 22 11.30 &   0.4 &   5.0 &   0.1 \\
HC435 & 5 35 16.20 & -5 22 37.50 &  -3.7 &   2.8 &  -1.3 \\
HC758 & 5 35 16.24 & -5 22 24.30 &   3.5 &   2.2 &   1.6 \\
HC317 & 5 35 16.24 & -5 23 19.10 &  -0.3 &   1.8 &  -0.1 \\
163-210 & 5 35 16.27 & -5 22 10.45 &   2.2 &   4.2 &   0.5 \\
HC522 & 5 35 16.29 & -5 22 10.40 &   2.2 &   4.1 &   0.5 \\
163-222 & 5 35 16.30 & -5 22 21.50 &   3.0 &   2.1 &   1.4 \\
HC479 & 5 35 16.31 & -5 22 24.00 &   1.4 &   2.0 &   0.7 \\
HC488 & 5 35 16.32 & -5 22 21.60 &   3.0 &   2.1 &   1.4 \\
163-249 & 5 35 16.33 & -5 22 49.01 &  -1.3 &   1.5 &  -0.9 \\
HC292 & 5 35 16.35 & -5 23 25.30 &  -0.2 &   1.8 &  -0.1 \\
HC485 & 5 35 16.38 & -5 22 22.30 &  -0.8 &   1.9 &  -0.4 \\
HC341 & 5 35 16.41 & -5 23 11.50 &   1.1 &   2.0 &   0.5 \\
HC514 & 5 35 16.43 & -5 22 12.20 &   6.7 &   2.7 &   2.4 \\
HC309 & 5 35 16.46 & -5 23 23.00 &  -1.3 &   1.6 &  -0.8 \\
165-235 & 5 35 16.48 & -5 22 35.16 &  -0.3 &   1.5 &  -0.2 \\
HC390 & 5 35 16.49 & -5 22 56.50 &   1.4 &   1.1 &   1.3 \\
HC442 & 5 35 16.50 & -5 22 35.20 &  -0.3 &   1.5 &  -0.2 \\
165-254 & 5 35 16.54 & -5 22 53.70 &  -0.1 &   1.2 &  -0.1 \\
166-250 & 5 35 16.59 & -5 22 50.36 &  -0.1 &   1.5 &  -0.0 \\
166-316 & 5 35 16.61 & -5 23 16.19 &   1.8 &   2.5 &   0.7 \\
HC325 & 5 35 16.63 & -5 23 16.10 &   3.6 &   2.4 &   1.5 \\
HC567 & 5 35 16.65 & -5 21 52.70 &  -0.4 &   5.1 &  -0.1 \\
HC280 & 5 35 16.66 & -5 23 28.90 &  -0.5 &   2.8 &  -0.2 \\
HC293 & 5 35 16.73 & -5 23 25.20 &  -0.2 &   2.3 &  -0.1 \\
167-231 & 5 35 16.73 & -5 22 31.30 &   3.0 &   1.6 &   1.9 \\
168-328 & 5 35 16.77 & -5 23 28.06 &   5.5 &   2.8 &   2.0 \\
HC518 & 5 35 16.78 & -5 22 11.70 &  -1.3 &   2.1 &  -0.6 \\
168-235 & 5 35 16.81 & -5 22 34.71 &   0.6 &   1.3 &   0.5 \\
HC349 & 5 35 16.87 & -5 23  7.10 &   0.4 &   0.7 &   0.5 \\
HC484 & 5 35 16.90 & -5 22 22.50 &  -1.0 &   1.5 &  -0.7 \\
HC441 & 5 35 16.91 & -5 22 35.20 &   0.6 &   1.3 &   0.4 \\
HC397 & 5 35 16.91 & -5 22 55.10 &   1.7 &   0.8 &   2.2 \\
HC494 & 5 35 16.94 & -5 22 20.70 &  -1.0 &   1.4 &  -0.7 \\
HC524 & 5 35 16.94 & -5 22  9.90 &  -1.6 &   1.4 &  -1.2 \\
HC450 & 5 35 17.01 & -5 22 33.10 &  -0.2 &   1.5 &  -0.1 \\
HC410 & 5 35 17.12 & -5 22 50.10 &   0.5 &   1.3 &   0.4 \\
HC517 & 5 35 17.13 & -5 22 11.90 &  -2.2 &   1.3 &  -1.7 \\
HC315 & 5 35 17.16 & -5 23 20.40 &   1.6 &   1.3 &   1.3 \\
HC330 & 5 35 17.24 & -5 23 16.60 &  -0.2 &   1.1 &  -0.2 \\
HC493 & 5 35 17.34 & -5 22 21.20 &   1.0 &   0.8 &   1.2 \\
174-305 & 5 35 17.37 & -5 23  4.86 &   0.7 &   0.5 &   1.3 \\
HC540 & 5 35 17.40 & -5 22  3.70 &   0.5 &   1.0 &   0.5 \\
HC313 & 5 35 17.47 & -5 23 21.10 &   1.8 &   1.3 &   1.4 \\
176-325 & 5 35 17.55 & -5 23 24.96 &   4.0 &   1.6 &   2.5 \\
HC550 & 5 35 17.55 & -5 22  0.30 &   0.5 &   1.1 &   0.4 \\
HC388 & 5 35 17.56 & -5 22 56.80 &   1.6 &   0.7 &   2.3 \\
HC295 & 5 35 17.57 & -5 23 24.90 &   4.0 &   1.6 &   2.5 \\
HC469 & 5 35 17.58 & -5 22 27.80 &   0.4 &   0.6 &   0.7 \\
HC513 & 5 35 17.62 & -5 22 12.60 &   2.2 &   0.7 &   3.4 \\
HC563 & 5 35 17.62 & -5 21 53.90 &   1.7 &   2.0 &   0.9 \\
176-252 & 5 35 17.64 & -5 22 51.66 &   0.3 &   0.7 &   0.4 \\
HC527 & 5 35 17.66 & -5 22  7.90 &   0.8 &   0.6 &   1.2 \\
HC405 & 5 35 17.66 & -5 22 51.70 &   0.6 &   0.7 &   0.9 \\
HC333 & 5 35 17.74 & -5 23 14.90 &   1.5 &   1.0 &   1.5 \\
HC462 & 5 35 17.76 & -5 22 31.00 &   1.8 &   0.6 &   3.0 \\
HC332 & 5 35 17.82 & -5 23 15.60 &   3.0 &   1.0 &   3.0 \\
HC496 & 5 35 17.83 & -5 22 19.60 &  -0.2 &   0.8 &  -0.2 \\
HC367 & 5 35 17.87 & -5 23  3.10 &   0.4 &   0.6 &   0.7 \\
HC542 & 5 35 17.88 & -5 22  3.00 &  -0.5 &   0.8 &  -0.6 \\
HC425 & 5 35 17.95 & -5 22 45.50 &   1.1 &   0.6 &   1.8 \\
HC501 & 5 35 18.03 & -5 22 18.20 &   2.6 &   0.8 &   3.4 \\
HC535 & 5 35 18.03 & -5 22  5.50 &  -0.6 &   0.6 &  -1.0 \\
180-331 & 5 35 18.03 & -5 23 30.80 &   4.2 &   1.2 &   3.5 \\
HC271 & 5 35 18.05 & -5 23 30.80 &   4.2 &   1.2 &   3.6 \\
HC372 & 5 35 18.08 & -5 23  1.80 &   0.3 &   0.6 &   0.5 \\
182-316 & 5 35 18.19 & -5 23 31.55 &   0.2 &   1.0 &   0.2 \\
HC533 & 5 35 18.24 & -5 22  6.30 &   0.4 &   0.4 &   0.9 \\
HC331 & 5 35 18.25 & -5 23 15.70 &   2.3 &   0.8 &   2.7 \\
HC348 & 5 35 18.28 & -5 23  7.50 &   0.3 &   0.4 &   0.8 \\
HC430 & 5 35 18.40 & -5 22 40.00 &   0.6 &   0.5 &   1.1 \\
HC278 & 5 35 18.50 & -5 23 29.30 &   0.2 &   0.9 &   0.2 \\
HC384 & 5 35 18.53 & -5 22 58.10 &   0.6 &   0.4 &   1.5 \\
HC463 & 5 35 18.58 & -5 22 31.00 &   1.1 &   0.5 &   2.4 \\
HC337 & 5 35 18.67 & -5 23 14.00 &   1.2 &   0.8 &   1.5 \\
HC713 & 5 35 18.71 & -5 22 56.90 &   1.0 &   0.4 &   2.6 \\
HC543 & 5 35 18.76 & -5 22  2.20 &   0.4 &   0.4 &   0.8 \\
189-329 & 5 35 18.87 & -5 23 28.85 &   2.6 &   0.9 &   2.8 \\
HC352 & 5 35 18.88 & -5 23  7.20 &   0.6 &   0.6 &   1.0 \\
HC498 & 5 35 18.96 & -5 22 18.80 &   0.3 &   0.8 &   0.4 \\
HC311 & 5 35 18.97 & -5 23 22.00 &   0.0 &   0.7 &   0.0 \\
190-251 & 5 35 19.03 & -5 22 50.65 &   1.0 &   0.4 &   2.4 \\
HC357 & 5 35 19.11 & -5 23  6.30 &  -0.4 &   0.6 &  -0.7 \\
HC288 & 5 35 19.12 & -5 23 27.10 &   1.4 &   0.8 &   1.8 \\
HC444 & 5 35 19.14 & -5 22 34.60 &  -0.1 &   0.4 &  -0.2 \\
HC408 & 5 35 19.22 & -5 22 50.70 &  -0.2 &   0.4 &  -0.6 \\
HC356 & 5 35 19.38 & -5 23  6.50 &   0.6 &   0.6 &   1.0 \\
HC491 & 5 35 19.47 & -5 22 21.80 &   1.8 &   0.6 &   2.9 \\
HC446 & 5 35 19.68 & -5 22 34.20 &   0.1 &   0.3 &   0.4 \\
HC531 & 5 35 19.90 & -5 22  7.30 &  -0.7 &   0.4 &  -1.8 \\
HC564 & 5 35 19.97 & -5 21 54.00 &   0.3 &   0.6 &   0.6 \\
HC452 & 5 35 19.98 & -5 22 32.80 &   0.3 &   0.6 &   0.5 \\
HC766 & 5 35 20.00 & -5 23 28.80 &   1.0 &   0.5 &   2.2 \\
HC474 & 5 35 20.03 & -5 22 26.50 &   0.2 &   0.7 &   0.3 \\
HC365 & 5 35 20.13 & -5 23  4.50 &   0.8 &   0.5 &   1.6 \\
HC346 & 5 35 20.18 & -5 23  8.50 &   0.6 &   0.6 &   1.0 \\
HC510 & 5 35 20.40 & -5 22 13.70 &  -0.0 &   0.5 &  -0.1 \\
\enddata
\end{deluxetable}

\end{document}